\documentclass[fleqn,10pt, table]{wlscirep}
\usepackage[utf8]{inputenc}
\usepackage[T1]{fontenc}
\usepackage{amsmath}

\usepackage{stackengine}

\usepackage{multirow}
\usepackage{xcolor}
\usepackage[most]{tcolorbox}

\newcommand{\metricboxup}[1]{\rotatebox[origin=c]{90}{\parbox{1.2cm}{\centering #1}}$\uparrow$}

\newcommand{\metricboxdown}[1]{\rotatebox[origin=c]{90}{\parbox{1.2cm}{\centering #1}}$\downarrow$}

\title{Similarity and Quality Metrics for MR Image-to-Image Translation}

\author[1]{Melanie Dohmen}
\author[1]{Mark A. Klemens}
\author[1]{Ivo M. Baltruschat}
\author[1]{Tuan Truong}
\author[1]{Matthias Lenga}
\affil[1]{\centering\stackunder{Bayer AG, Radiology, Berlin, Germany}{\emph{firstname.lastname@bayer.com}}}
\begin{abstract}
Image-to-image translation can create large impact in medical imaging, as images can be synthetically transformed to other modalities, sequence types, higher resolutions or lower noise levels.
To ensure patient safety, these methods should be validated by human readers, which requires a considerable amount of time and costs. Quantitative metrics can effectively complement such studies and provide reproducible and objective assessment of synthetic images. 
If a reference is available, the similarity of MR images is frequently evaluated by SSIM and PSNR metrics, even though these metrics are not or too sensitive regarding specific distortions. When reference images to compare with are not available, non-reference quality metrics can reliably detect specific distortions, such as blurriness.
To provide an overview on distortion sensitivity, we quantitatively analyze 11 similarity (reference) and 12 quality (non-reference) metrics for assessing synthetic images. We additionally include a metric on a downstream segmentation task. We investigate the sensitivity regarding 11 kinds of distortions and typical MR artifacts, and analyze the influence of different normalization methods on each metric and distortion. Finally, we derive recommendations for effective usage of the analyzed similarity and quality metrics for evaluation of image-to-image translation models.
\end{abstract}


\begin{document}

\flushbottom
\maketitle

\thispagestyle{empty}

\section{Introduction}
\subsection{Image Synthesis}
Recent advances in generative artificial intelligence (AI) within the natural image domain have demonstrated a remarkable capability to produce synthetic images with high fidelity, capturing nuances such as lighting variations, textures, and object placements\cite{palette, pix2pix}. The implications of these advancements are far-reaching, with applications spanning various domains such as computer vision, graphics, augmented and virtual reality, or creative arts. Still, many challenges remain, including potential biases in generated images, the need for enhanced diversity and controllability in generation, and ethical considerations surrounding the use of AI-generated content. 
\begin{figure}[htb]
\centering
\includegraphics{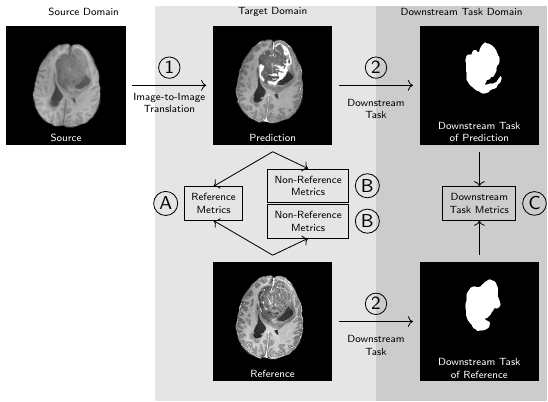}
\caption{Overview of image-to-image translation and types of evaluation metrics. (1) A source image from a source domain is transformed to a prediction in the target domain by an image-to-image translation model. If a reference image is given, this also belongs to the target domain. Then there are multiple possibilities to apply metrics. (A) Reference metrics directly compare prediction and reference image. (B) Non-reference metrics can be applied to the prediction alone, but also - if available - to a reference image. Then both non-reference metric scores can be compared. As an additional option (C), the reference and the prediction can be further processed in a downstream task, i.e. a segmentation task as a second (2) step. The performance of both downstream task results is then assessed with a downstream task metric, i.e. a segmentation metric.}
\label{fig:overview}
\end{figure}
The adaptation of generative modeling concepts such as Generative Adversarial Networks (GANs)\cite{Review_GANs_medical} or diffusion models \cite{diffusion_models, latent_diffusion_medical} to the medical imaging domain is being explored with continuously growing interest and many relevant use cases have already been identified, such as data augmentation \cite{data_augment} or image conversion and enhancement by image-to-image translation \cite{image2image_appl_medical}.
For training deep neural networks, typically very large and diverse data sets are needed, but these are rarely available for medical imaging tasks. Generative networks can amend available data with synthetic samples and thereby improve the performance of other image analysis tasks \cite{data_augment}.

\subsection{Image-to-Image Translation}
Another line of research develops conditional generative models for image-to-image translation, which aims to translate a given source image to a synthetic target image, showing the same content (e.g. patient, organ or biological sample) as the source but with a different appearance (e.g. contrasting structures differently, changing texture or resolution). Source and target typically belong to different image domains.
Depending on the availability of source and target pairs of the same patient or structure in the training data, image-to-image translation can be performed in a paired \cite{pix2pix} or unpaired \cite{cycleGAN} manner, also referred to as supervised or unsupervised. Image-to-image translation models are trained to transfer characteristics from the target domain to a specific image from the source domain without changing the represented image content.

\subsection{Medical Image Synthesis Tasks}
These tasks are specifically interesting in medical applications \cite{image2image_appl_medical}, because they allow to translate a medical image from one domain to another. Source and target domain may differ by imaging modality or acquisition parameters, such that image-to-image translation allows translation, e.g., from computed tomography (CT) to magnetic resonance (MR) imaging \cite{LumbarCT_to_MR} or vice-versa \cite{MR2CT}, from T1-weighted MR to T2-weighted MR \cite{Make-A-Volume_CrossMRISynthesis}, CT to positron emission tomography (PET) \cite{virtualPET_from_CT}, PET to CT \cite{PET2CT}, or from native MR to contrast-enhanced MR \cite{ammari2022,baltruschat2023freggan}. There is a significant patient benefit, when the source image can be acquired with less harm, more quickly or at a lower cost, compared to the the target image, which might be preferred for diagnosis.
Low-quality or low-resolution images can be restored, improved, or the resolution increased \cite{CARE}.
For radiation therapy, which is planned on the basis of CT images, MR to CT synthesis has been investigated \cite{MR2CT}. MR to CT synthesis is also used to facilitate registration between both modalities\cite{syn_for_registration}.
The translation between different MR sequences (T1-weighted, T2-weighted, T2-FLAIR) can complete missing series for improved diagnosis \cite{raut2024, MissingMRISequence, Multi-ModalMRISynthesis, mustGAN}. 
For reducing patient burden with contrast agents, researchers work on the translation of native or low-dose MR images to synthetic high contrast-enhanced images \cite{mallio2023}.

\subsection{Validation Metrics for Image Synthesis}
%
However, validation of these approaches is not straightforward. If a reference image, representing the desired synthesis result, is available for each generated image, a group of metrics called reference metrics can assess the similarity between predicted and reference images (see Fig.\,\ref{fig:overview} (A)). These reference images are often already leveraged for training synthesis models in a paired or supervised manner. Reference metrics are sometimes called full-reference metric to distinguish from weak-reference metrics, that only use partial information or features of the reference image. The term similarity metric is also synonymously used for reference metric.

%
However, not always paired reference images are available. In this case, non-reference metrics, also called quality metrics, can be applied (see Fig.\,\ref{fig:overview} (B)). As quality requirements may vary between tasks and image domains, metrics for different aspects have been developed, e.g. for measuring blurriness \cite{blur_effect}, contrast  \cite{MichelsonContrast}, noisiness \cite{total_variation} or other features inspired by human perception \cite{pianykh2018}. 
%
Depending on the application of synthesized images, the evaluation of the images in a downstream task is more appropriate, than the evaluation of the images themselves \cite{BeautyOrBeast}. Downstream task metrics operate on further processed results and not on the predicted or reference image (see  Fig.\,\ref{fig:overview} (C)).
%
For GANs, so-called distribution based metrics are very popular \cite{GAN_evaluation}. These assess the distributions of extracted image features of a larger set of generated images. For example, the Inception score \cite{ImprovedGANTraining} assesses how distinct and evenly distributed classes are predicted by an Inception architecture based classifier trained on ImageNet (InceptionNet \cite{Inception}). And the Frêchet Inception Distance\cite{FID} assesses how well Gaussian modeled activation layer distributions of the InceptionNet match between generated and reference image sets. As these metrics do not assess single images, these metrics are not in the scope of this study.

In the domain of natural images, reference metrics have been extensively tested on synthetically distorted images.
The Tampere Image Database \cite{tid_dataset} and the LIVE Image Quality Assessment Database \cite{full_reference_metrics, LIVE} contain 25-29 reference images and differently distorted versions thereof, additionally annotated with a human quality score. Metric scores for various full-reference metrics were correlated with human scores to identify the best performing metrics.

Specifically for image synthesis, a study \cite{ding2020optim} assessed metrics on outputs of image synthesis results.
However, these results cannot be fully transferred to medical image synthesis. Even though similar studies for medical images \cite{SubjectiveAssessmentMR, kastryulin} exist, the included distortions, such as JPEG compression artifacts and white noise are less relevant for medical image synthesis. Instead, MR imaging, including acquisition and reconstruction, exhibits very specific artifacts, such as bias field, ghosting or stripe artifacts. Additionally, certain synthesis models may introduce other kinds of distortions, e.g. the insertion of artificial structures or registration artifacts that arise from misaligned source and target images. In this study, we create a similar benchmark dataset for the medical images domain consisting of 100 MR reference images and 11 mostly MR specific distortions. Applying the distortions in an isolated manner in five defined strengths results in well defined distortions. This allows us to compare different metrics regarding their sensitivity towards each kind of distortion separately, instead of averaging over a fixed group of distortions. This is crucial, because the effect of distortions on image quality may be rated differently for different medical applications.
\subsection{Validation in the Medical Domain}
Analysis, processing and generation of medical images can have severe impact on patient outcome and patient safety. Therefore, the Food and Drug Administration (FDA) in the United States requires technical and clinical evaluation for every software as a medical device to be approved\cite{FDA_guidelines}. 
Technical validation provides objective evidence, that the software correctly and reliably processes input data and generates output data with the appropriate level of accuracy and reproducibility. Clinical validation measures the ability of a software to yield a clinically meaningful output in the target health care situation. For evaluating medical algorithms based on artificial intelligence, guidelines about trial protocol designs \cite{SPIRIT-AI} have been agreed on by an expert consortium. However, systemic reviews of published papers on AI-based algorithms in the medical domain have revealed, that only a small fraction of studies adheres to such guidelines, i.e. external validation\cite{Design_AI_Studies, validation_against_HCP}. Often, details of clinical validation studies, such as test population statistics, are not published, not even for FDA-approved software\cite{Trends_FDA_approval}. Another review found that the median number of health-care professionals engaged in clinical validation was only four \cite{AI_vs_Clinicians}, which limits the reliability and generizability of such studies. Recent validation studies of FDA approved image-to-image translation software rely strongly on technical assessment of phantom measurements by similarity metrics \cite{validation_CT_recon, validation_SubtlePET}. While the number of FDA approved medical devices based on AI software is still increasing by 14\% (2022), the increase has been slowing down compared to 2020 (39\%) \cite{fda}.

The lack of adequate trials for clinical validation is certainly only one part of the problem. Appropriate technical validation is crucial at an even earlier stage of development. Metrics for  biomedical image analysis and segmentation have been extensively described \cite{metric_pitfalls, metrics_reloaded}, and can be leveraged to indirectly assess synthetic images via downstream tasks \cite{baltruschat2024brasyn}. Even though a huge amount of metrics has been used for the evaluation of synthetic medical images \cite{Necasova2022, Review_GANs_medical}, to our knowledge guidelines for the selection of appropriate similarity and quality metrics are not available.
Loss functions for medical image registration, if used to compare a synthetic image with its reference image, are also leveraged for measuring image similarity, and overlap strongly with the selected reference metrics in this study \cite{survey_registration_metrics}.
For validating synthetic MR images, especially structural similarity index measure (SSIM), and peak signal-to-noise ratio (PSNR) are used extensively. A review on image-to-image translation with generative adversarial networks (GANs) or convolutional neural networks (CNNs) in the medical domain \cite{MacNaughton2023} reported the use of SSIM in 84\% studies and PSNR in 61\% of studies, that synthesize MR images. A further review on synthetic contrast-enhancement of MR images \cite{haase2023_review} reports evaluation by SSIM and PSNR in 75\% of the studies. This is in opposition to known crucial weak points of SSIM and PSNR, such as underestimated blurriness 
\cite{Mudeng2022}, bad correlation to human perception \cite{psnr_perception, psnr_revisited, MSE_loveit_or_leaveit} and difficulties with float-valued images \cite{SSIM_properUse, floating_point_SSIM}. Therefore, a systematic analysis of appropriate metrics for MR image synthesis validation is needed.
\subsection{Contribution}
In this paper, we provide a comprehensive analysis of the sensitivity of 11 reference and 12 non-reference metrics to 11 different distortions, that are relevant for MR image synthesis and of which some have not been assessed with the selected metrics before.
Furthermore, we investigate the influence of five normalization methods before metric assessment.
After analyzing metric sensitivity in detail, and discussing specific shortcomings or advantages of the investigated metrics, we recommend how to select and best apply metrics for validating image-to-image translation methods specifically for MR image synthesis.

\section{Methods}
\label{sec:methods}
In this section, we give an overview of reference and non-reference metrics for assessing the quality of images. Since most reference and non-reference metrics strongly depend on the intensity value ranges of the images they assess, the examination of metrics must be combined with the examination of normalization methods, that adjust the intensity value ranges. Therefore, we first introduce normalization methods that are frequently used to bring MR images to a common intensity value scale or as prerequisite for certain metrics.
\subsection{Intensity Ranges and Data Formats} \label{sec:data_ranges}
For two reasons, normalization of medical images is needed prior to similarity or quality assessment.
First, image intensities of non-quantitative image modalities are not comparable between two images, due to missing standardization. For example, in MR imaging, the same tissue may be represented by different values depending on scanner, software version or surrounding tissue. In this case, normalization must be applied in order to achieve comparability between two or more images.
Normalization or standardization is not only needed as a prerequisite for metric assessment but is usually already performed as a preprocessing step for image-to-image translation models. For deep-learning based methods, a reasonable and standardized scale such as $[-1, 1]$ or $[0, 1]$ is recommended \cite{preprocessing_DL_book}.  
By modality specific normalization, deep-learning based models may even improve generalizability in case of heterogeneous input data sources \cite{stain_normalization, MR_normalization_multisite}.

Second, most metrics were designed and developed for 8-bit unsigned integer data format.
 In many cases, medical images are acquired in a larger 16-bit integer or 32-bit float value range and need to be normalized into the $[0,255]$ range. Alternatively, an additional data range parameter is introduced to adapt to other intensity ranges. For 8-bit images, the data range parameter $L$ is then set to $255$, assuming an intensity value range between $0$ and $255$. For float valued images, the intensity value range is generally infinite, but $L$ should be set to a finite value spanning the range of at least all observed intensity values. Therefore, $L$ is typically set to the difference of maximum and minimum value of an image $I$ ($L_I = I_{\mathrm{max}} - I_{\mathrm{min}}$), or the difference of the joint maximum and minimum of two images $I$ and $R$ ($L_{I,R}$), or of a set of images $\mathcal{D}$ ($L_{\mathcal{D}}$). If $I, R \in \mathcal{D}$, then $L_I \le L_{I,R} \le L_{\mathcal{D}}$. It is argued \cite{SSIM_properUse}, that using $L_{\mathcal{D}}$ results in SSIM values, that do not vary with individual image minimum and maximum values. At least $L_{I,R}$ should be used for reference metrics on two images $I$ and $R$ instead of $L_I$ or $L_R$, because possibly $L_I \ne L_R$, and then the reference metric would yield different results when interchanging $I$ and $R$.
\subsection{Normalization Methods}
\label{sec:normalization_methods}
Several normalization methods\cite{MR_normalization, MR_normalization_multisite}, such as Zscore,  Minmax or Quantile normalization have been used for MR images. These normalization methods ensure, that intensity values are near $[-1, 1]$ or strictly between $[0, 1]$, before model training \cite{ammari2022, Haase2023}.
Other variants of Minmax utilize percentiles or quantiles for scaling and/or additional clipping, or estimate percentiles based on a region-of-interest (ROI) contained in the image. For example, the WhiteStripe normalization \cite{WhiteStripe} determines reference ranges in a white matter region of the brain. Afterwards, these parameters are used to shift and scale the intensity values.

In addition, normalization methods have been developed specifically for MR (piece-wise linear histogram matching \cite{nyul1999}) or even specifically for brain MR images (cf. WhiteStripe \cite{WhiteStripe}) in order to obtain quantitative comparable intensities for the same brain or body structures. 
\begin{table}[htb]
\centering
\caption{Overview of normalization methods. The target intensity range $[j_1, j_2]$ can be chosen arbitrarily, but is typically set to $[0, 1]$ or $[-1, 1]$. }
\label{tab:normalization}
\begin{tabular}{l|p{8cm}|p{4.5cm}|l}
\toprule
Method & Description & Parameters & References \\
\midrule
Minmax & Shifts and scales image intensity values to a range with given minimum and maximum value. & result range: $[j_1, j_2]$ & \cite{MR_normalization_multisite}\\
cMinmax & Clips at lower and upper percentiles before Minmax. More robust to high and low outliers.& result range: $[j_1, j_2]$\\
ZScore &  Shifts the intensity values to a mean $\mu_I$ of 0 and scales to unit standard deviation $\sigma_I$. &  $\mu_I=0, \sigma_I=1$ & \cite{MR_normalization_multisite} \\
Quantile & Shifts the intensity values to a median of 0 and scales to a unit inter-quartile range (IQR = $I_{p_{75\%}} - I_{p_{25\%}}$). &  median($I$)=0, IQR=1 & \cite{MR_normalization_multisite} \\
Binning & Binning: All intensity values are mapped to B (=256) equidistant bins. & \parbox[t]{3cm}{\centering discrete result range:\\$[0, B-1]$}  & \\
PL & Piecewise-Linear: The histogram is scaled linearly in two pieces to match three landmarks $s_1$, $m_s$, and $s_2$ of a standard histogram derived from a set of reference images. &\parbox[t]{3cm}{\centering result range: $[s_1, s_2]$,\\ mode: $m_s$, depends on a training dataset}&\cite{nyul1999}\\
\bottomrule
\end{tabular}
\end{table}
In a similar fashion, other normalization types besides Zscore can be adapted. For instance, for deep learning based MRI liver tumor segmentation \cite{Haensch2022}, a Minmax normalization was applied by using the 2\% and 98\% percentiles $I^{\text{Liver}}_{2\%}$ and $I^{\text{Liver}}_{98\%}$ of the local intensity distribution within a region of interest in the liver for rescaling.
In order to match histogram modes and minimum and maximum percentiles of an MR dataset showing the same body region, a piece-wise linear standardization procedure\cite{nyul1999} has been proposed. The authors argue that MR images exhibit an unimodal or bimodal histogram for most body regions, where the foreground concentrates around the first or second mode.

However, we assume, that different MRI sequences may generally display different histogram shapes and a different distribution of contrasted tissue types. Therefore, each MRI sequence should be normalized separately. Possibly, detecting and excluding the background could be beneficially done simultaneously for multi-modal MR images.
Binning can be regarded as a normalization method to convert float values to 8-bit integer values. In this case 256 bins are used. It also removes information, because close but different intensities are mapped to the same bin value. By additionally copying the binned gray value to three color channels, images can further be converted to RGB images. Binning is needed for some metrics, because they require a finite number of intensity values (see Sec.\ref{sec:nmi}), or it is used to easily apply a method for 8-bit images to a binned float valued image.
The normalization methods investigated within this study are defined in detail in Supplementary Sec.\,A.2, an overview is shown in Tab.\ref{tab:normalization}.
\subsection{Reference Metrics}
\label{sec:ref_metrics}
\begin{table}[thb]
\centering
\caption{Overview of reference (similarity) and non-reference (quality) metrics. The arrows indicate if the metric increases ($\uparrow$) or decreases ($\downarrow$) with increasing similarity or quality. [Implementation sources: gh=gitHub, itk=Insight Segmentation and Registration Toolkit\cite{itk}, np=numpy\cite{numpy}, pypi=python package index, skl=scikit-learn\cite{scikit-learn}, ski=scikit-image\cite{skimage}, tm=torchmetrics\cite{torchmetrics}]}
\begin{tabular}{c |p{0.6cm}|p{1.5cm}|p{6.5cm}|c|c|l}
\multicolumn{2}{c|}{Group} & \parbox{1cm}{\centering Abbre-viation}& Description & \parbox{1.7cm}{\centering Similarity $\uparrow$\newline [min, max]} & \parbox{1cm}{\centering Imple-mentation} & Ref. \\
\toprule
\multirow[c]{21}{*}{\rotatebox[origin=c]{90}{ Reference (Similarity) Metrics}} &\multirow[c]{7}{*}{\rotatebox[origin=c]{90}{SSIM}}& SSIM & Structural Similarity Index Measure: combination of structure, luminance and contrast & $[0, 1] \uparrow$ & tm, ski& \cite{ssim} \\
&& MS-SSIM & Multi-Scale SSIM: SSIM on original and 4 downscaled image resolutions& $[0, 1] \uparrow$ & tm & \cite{msssim}\\
&& CW-SSIM & Complex Wavelet SSIM: ignores phase shifts in the wavelet domain, ignores small rotations and spatial translations & $[0, 1] \uparrow$ & gh \cite{Ding_code} &\cite{cw-ssim} \\
\cline{2-7}
&\multirow[c]{2}{*}{\rotatebox[origin=c]{90}{PSNR}}& PSNR & Peak Signal-to-Noise-Ratio: relation of data range to MSE& $[0, \infty] \uparrow$ & tm, ski & \cite{psnr}\\
\cline{2-7}
&\multirow[c]{3}{*}{\rotatebox[origin=c]{90}{\parbox{1.1cm}{\centering Error Metrics}}}& NMSE & Normalized Mean Squared Error & $[0, \infty] \downarrow$ & \\
&& MSE & Mean Squared Error & $[0, \infty] \downarrow$ & skl \\
&& MAE & Mean Absolute Error & $[0, \infty] \downarrow$ &  skl\\
\cline{2-7}
&\multirow[c]{3}{*}{\rotatebox[origin=c]{90}{\parbox{1.2cm}{\centering Learned Metrics}}}
& LPIPS & Learned Perceptual Image Patch Similarity & $[0, 1] \downarrow$ & pypi \cite{lpips_package}, tm & \cite{lpips}\\
&& DISTS & Deep Image Structure and Texture Similarity Metric  & $[0, 1] \downarrow$ & gh\cite{DISTS_code, Ding_code} & \cite{DISTS}\\
\cline{2-7}
&\multirow[c]{3}{*}{\rotatebox[origin=c]{90}{\parbox{1.4cm}{\centering Statist. Depend.}}}& NMI & Normalized Mutual Information: MI with fixed range & $[1, 2] \uparrow$ & ski & \\
&& PCC & Pearson Correlation Coefficient & 
$[0,1] \uparrow$ & skl &\\ 
\cline{1-7}
&\multirow[c]{3}{*}{\rotatebox[origin=c]{90}{\parbox{1cm}{Down-stream}}}& DSC & Dice Similarity Coefficient: segmentation metric, evaluating overlap & $[0,1] \uparrow$ & itk & \\
&&&&&&\\
\hline
\hline
\multicolumn{4}{c}{}& Quality $\uparrow$ &\multicolumn{2}{c}{} \\
\cline{2-7}
\multirow[c]{23}{*}{\rotatebox[origin=c]{90}{Non-Reference (Quality) Metrics}}&
\multirow[c]{10}{*}{\rotatebox[origin=c]{90}{Blurriness}}& BE & Blur Effect: difference of gradients when additionally blurred & $\downarrow$ &ski & \cite{blur_effect}\\ 
&& BR & Blur Ratio: ratio of blurred pixels to edge pixels  & $\downarrow$  & - & \cite{blur2}\\
&& MB & Mean Blur: sum of inverse blurriness divided by number of blurred pixels & $\uparrow$  & - & \cite{blur2}\\
&& VL & Variance of Laplacian & $\downarrow$ & ski+np  & \cite{VarLapl}\\
&& BEW & Blurred Edge Widths & $\downarrow$ & - & \cite{BlurMarz}\\
&& JNB & Just Noticeable Blur & $\downarrow$ & gh(C++)\cite{JNB_implementation} & \cite{BlurJNB}\\
&& CPBD & Cumulative Probability of Blur Detection & $\uparrow$ & gh\cite{CPBD_implementation} & \cite{BlurCPBD}\\
\cline{2-7}
&\multirow[c]{7}{*}{\rotatebox[origin=c]{90}{MR Quality}}& MLC & Mean Line Correlation (also average structural noise): mean correlation between neighbored rows and columns & $\downarrow$ & - & \cite{schuppert2022}\\
&& MSLC & Mean Shifted Line Correlation (also average nyquist ghosting): mean correlation between rows and columns, that are with half image distance apart & $\downarrow$ & -&\cite{schuppert2022}\\
\cline{2-7}
&\multirow[c]{3}{*}{\rotatebox[origin=c]{90}{\parbox{1.3cm}{\centering Learned Quality}}}& BRISQUE & Blind/ Referenceless Image Spatial Quality Evaluator & $\downarrow$ & pypi\cite{pypi_brisque} & \cite{brisque}\\
&& NIQE & Natural Image Quality Evaluator & $\downarrow$ & gh\cite{niqe_implementation}&\cite{niqe}\\
\cline{2-7}
&\multirow[c]{3}{*}{\rotatebox[origin=c]{90}{\parbox{0.5cm}{\centering Noisi-ness}}}& MTV & Mean Total Variation: Mean L2-normed gradient in x- and y- direction & $\downarrow$ & -&\cite{total_variation}\\
\bottomrule
\end{tabular}
\label{tab:metrics} \label{tab:nr_metrics}
\end{table}
Reference metrics are based on comparing a reference image $R$ with another image $I$. Both images are assumed to have the same spatial dimensions. An overview of the reference metrics analyzed in this study is given in Tab.\,\ref{tab:metrics}. If the image $I$ was not acquired with the same modality or the same time point as image $R$, spatial alignment has to be ensured before applying a reference metric. Typically, this is achieved by image registration techniques.

Slight spatial misalignment between paired images has been identified as a problem when evaluating with reference metrics, such that specialized methods have been investigated for this purpose. By assessing similarity in the complex-wavelet domain, complex-wavelet SSIM (CW-SSIM) is able to ignore small translations, scaling and rotations \cite{cw-ssim}. A score derived from features of the Segment Anything Model (SAM), mainly compares semantic features and therefore better ignores different style and small deformations \cite{samscore}. The learned Deep Image Structure and Texture Similarity (DISTS) metric gives more weight to texture similarity, ignoring fine-grained misalignment of these textures \cite{DISTS}.

For natural images, a large set of reference metrics has been benchmarked on the Tampere Image Dataset \cite{tid_dataset} or the LIVE Image Quality Assessment Database \cite{LIVE}. Many of the standard metrics from natural imaging have been frequently applied to medical images. However, some careful modifications are necessary for images with a data type other than 8-bit unsigned integer. In the following, we introduce the investigated reference metrics with some more detail and background and highlight important adaptions. A full list of metrics and their calculation is found in Supplementary Sec.\,A.3.
\subsubsection{Structural Similarity Index Measure}
\label{sec:ssim}
The structural similarity index measure (SSIM) combines image structure, luminance, and contrast, which are calculated locally for each pixel \cite{ssim}.
Several variants of SSIM exist. Multi-scale SSIM (MS-SSIM) \cite{msssim} calculates local luminance, contrast and structure additionally for four downscaled versions of the images and combines them in a weighted fashion. MS-SSIM is more sensitive towards large-scale differences between the images to be compared. This puts less impact on high resolution details.
The complex-wavelet SSIM (CW-SSIM) \cite{cw-ssim} was specifically designed to compensate for small rotations and spatial translations. For this metric, only the coefficients of the complex-wavelet transformed images $I$ and $R$ are used, such that small phase shifts between both images are ignored. This additional freedom may allow unnatural results, when CW-SSIM is used for model optimization \cite{ding2020optim}.
As an adaption of SSIM to float valued images, the data range parameter $L$ (see Sec.\,\ref{sec:data_ranges}) is used to scale the internal constants $C_1 = (k_1\cdot L)^2 = (0.01 \cdot L)^2$ and $C_2 = (k_2\cdot L)^2 = (0.03 \cdot L)^1$ (and $C_3 = C_2/2$), which are used to make computations numerically stable (see supplement Sec.\,A.3, Eq.\,(41)). 
It can be derived from the SSIM calculation, that a high data range parameter $L$ and thereby high values for $C_1-C_3$, lead to SSIM values near 1, because the constants $C_1-C_3$ dominate the calculation compared to the observed intensity values. In these cases, SSIM is not very informative. This has been experimentally observed before \cite{floating_point_SSIM},
 and a default normalization of float value ranged images to the range $[0,1]$ and a modification of constants $C_1 = C_2$ to $(k_1 \cdot L)^2 = (0.0001\cdot 1)^2 = 1\cdot 10 ^{-8}$ was proposed.
Common implementations also define different default values for $L$. As the \texttt{skimage} \cite{skimage} package generally assumes float valued images to be in the range $[-1.0, 1.0]$, its implementation of SSIM defines $L=2$ as default. The \texttt{torchmetrics} \cite{torchmetrics} implementation of SSIM sets $L = L_{I, R}$ as default for images $I$ and $R$.

The choice for the \textbf{data range parameter $L$} is directly related to normalization techniques that rescale image intensities: If images $I$ and $R$ are scaled by factor $a$ to $I^{\prime} = I \cdot a$ and $R^{\prime} = R \cdot a$, then calculating SSIM on $I^{\prime}$ and $R^{\prime}$ with $L^{\prime} = L \cdot a$ will be the same as calculating SSIM on $I$ and $R$ with $L$. However, if image intensities are additionally shifted by an additive value $b$, the luminance term will increase with $b$ and yield different SSIM values. If this shift $b$ is negative, as it typically is with Zscore normalization (see Supplementary Sec.\,A.2, Eq.\,(3)), this can lead to negative SSIM values \cite{SSIM_properUse}. 
\subsubsection{Peak Signal-to-Noise Ratio}
Peak signal-to-noise ratio (PSNR) was developed to measure the reconstruction quality of a lossy compressed image compared to the uncompressed reference image \cite{psnr}. However, it is frequently used as a metric for assessing image similarity. The PSNR is infinite for identical images and decreases monotonically as the differences between image $I$ and reference $R$ increase. The data range parameter $L$ is incorporated in the PSNR as the peak signal. The noise in the PSNR is calculated as the mean squared error (MSE, see \ref{sec:MSE}).
For natural images, improved variants of PSNR called PSNR-HVS and PSNR-HVS-M have been developed, that seem to correlate closer to the human visual system \cite{tid_dataset}. Adapted implementations for variable intensity ranges and experiments with medical images are not available at this time. A deeper exploration of these metrics remains important future work.

\subsubsection{Error Metrics} \label{sec:MSE}
This group of metrics, including mean absolute error (MAE), mean square error (MSE), root mean square error (RMSE) and normalized mean square error (NMSE), directly depends on the absolute difference of intensity values at equal pixel locations.
The metrics MSE, RMSE, and NMSE are based on the squared difference and due to convex shape of the quadratic function, these metrics give more weight to large differences than MAE, which is based on the unsquared absolute difference. By normalization with the standard deviation of the reference image, NMSE assigns a higher similarity to images with a higher standard deviation, i.\,e. with high variation and a large range of intensity values. On the contrary, the same intensity differences lead to a lower similarity, if the reference image has a very low standard deviation, i.\,e. it appears very homogeneous. However, the scale and range of all these metrics strongly depend on the intensity value ranges and, thereby, also on the normalization method.
\subsubsection{Learned Metrics: LPIPS and DISTS}
Learned perceptual image patch similarity (LPIPS) relies on image feature maps from a trained image classification model. 
For the LPIPS metric, an Alex-Net or VGG-architecture backbone exist \cite{lpips}. The VGG version of LPIPS was recommended for usage as a traditional perceptual loss, while the Alex-Net version should be preferred as a forward metric. The latter one is also faster at inference due to the smaller network, so we analyzed LPIPS with Alex-Net in Sec. \ref{sec:distortion_experiments}. Even though the networks were trained on RGB images, the trained networks expect an input range of $[-1,1]$, e.g. by previous Minmax normalization.
LPIPS has shown great correlation with human perception and outperforms many other similarity metrics on natural images while the type of employed architecture has only minor influence \cite{lpips}. It has occasionally been used for validation of medical image synthesis \cite{OneModelToSynthesizeThemAll}, and is commonly applied as perceptual loss for training medical image-to-image translation models \cite{Review_GANs_medical}.

The Deep Image Structure and Texture Similarity (DISTS) metric is an adaption of LPIPS giving more focus on texture\cite{DISTS}.

\subsubsection{Statistical Dependency Metrics: NMI and PCC}
\label{sec:nmi}
Mutual information (MI) estimates the amount of information of an image $R$, that can be predicted from image $I$. MI is widely used as an optimization criterion for multi-modal image registration \cite{mutual_info}. It has been used sporadically as a metric for validation of image synthesis \cite{MR_normalization, MacNaughton2023}. 
The NMI has a fixed value range of $[1,2]$, which is preferable for comparing absolute metric scores and interpretability.
The Pearson correlation coefficient (PCC), is a statistical dependency metric which measures the degree of linear dependency between the intensities in $I$ and $R$ at each pixel location.
As PCC is defined by correlation and NMI and MI both operate on normalized binned images, previous normalization that purely scales and shifts, such as Minmax, Zscore or Quantile normalization, does not have any effect on the resulting scores.
The reference metrics investigated within this study are defined in detail in Supplementary Sec.\,A.3, an overview is shown in Tab.\,\ref{tab:metrics}.
\subsection{Indirect Evaluation with Downstream Tasks} \label{sec:seg_metrics}
Another option for validation is to consider which tasks are going to be performed downstream from a synthesized image. Whenever one of these tasks is performed, the quality of synthetic images can also be assessed by measuring the performance of the specific task on the image compared to the performance of the reference image.
In medical image-to-image translation, which aims for improved medical diagnosis or treatment, assessing the performance of medical diagnosis or treatment directly derived from digital images is very desirable. In this context, synthetic images must be processed in the same way as the reference image and should have equal outcomes in the downstream task. However, deviations between synthetic images and reference images can be accepted, when they have no impact on the downstream task.
As an example, if a synthetic MR image is generated for detecting a brain tumor, it is to some extent irrelevant if healthy brain tissue in the synthetic image appears slightly different than in the true reference image, as long as it is clearly identified as the same type of healthy tissue. If a synthetic histology image is rated with the same grade of cancer as the reference image, the exact cell-wise correspondence might not be important.
Many downstream tasks on medical images can be nowadays performed automatically, including:
\begin{itemize}
    \item Detection or segmentation of organs, cells and lesions, e.g. the Segmentation of brain tumors from T1-weighted native, T1-weighted contrast enhanced, T2-weighted, and fluid attenuation inversion recovery (FLAIR) MR images \cite{baltruschat2024brasyn} [Related downstream task metrics: DSC, Intersection over Union (IoU)].
    \item Classification of images or image segments, e.g. the synthesis of clinical skin images with 26 types of conditions, verified by classification scores of dermatologists \cite{DERMGAN} [Related downstream task metrics: Accuracy, Precision, Recall, F1-Score].
    \item Transfer learning and data augmentation, e.g. the synthesis of chest X-ray for data augmentation and evaluation of classification model on real data with and without synthetic training data \cite{synChestXray} [Related downstream task metrics: e.g. Sensitivity, Specificity, Area Under the Receiver-Operator-Characteristic curve (AUROC)].
 \item Multi-modal registration, e.g. the registration of synthesized MR image from CT image to MRI atlas instead of registration of CT image to MRI atlas\cite{CT_MRI_Registration} [Related downstream task metrics: MSE, MI].
 \item Dose calculation in radiation therapy planning, e.g. the synthesis of a planning CT from MRI for use in a radiation planning tool \cite{Dosimetric} [Related downstream task metrics: relative difference of planned radiation dose].
\end{itemize}
Detection, segmentation and classification metrics for the biomedical domain have been well documented and discussed \cite{metric_pitfalls, metrics_reloaded}. Also, a study of a segmentation metric systematically analyzed the sensitivity to relevant simulated distortions \cite{multiregionDSC}.
Therefore, the performance of such tasks with synthetic images can be well compared to the performance with reference images to validate the use of synthetic images for a specific task. The concept of downstream task evaluation metrics recognizes that the final goal of image synthesis in the medical domain is to generate useful and correct images rather than images, that are visually appealing \cite{TheBeautyAndTheBeast}. 
However, if image synthesis was optimized regarding a certain downstream task, the resulting images might not be optimal for other non-related tasks. Specifically, they might have a fake appearance, that does not interfere with the downstream task, but would be misleading for direct review of medical practitioners. Furthermore, the evaluation of downstream tasks can substantially depend on the performance of the downstream task method. If a segmentation model fails on a large set of reference images, the comparison to segmentations on synthetic images is obsolete.
The amount and variety of downstream tasks and corresponding metrics are almost unlimited, but to discuss and analyze the value of downstream tasks, we include the evaluation of a downstream segmentation model with a popular segmentation metric, namely the Dice Similarity Coefficient (DSC) \cite{dice_original, metric_pitfalls}. 
\subsection{Non-Reference Quality Metrics}
\label{sec:nr_metrics}
Non-reference metrics, often also called quality metrics or blind metrics, try to assess the quality of a distorted image without knowing the undistorted reference. As a reference might not be available, these metrics can be applied in many evaluation settings. However, there is a huge amount of such metrics and most of them assume a certain kind of distortion to be detected. The correlation of many of these metrics with human perception has been investigated \cite{pianykh2018}. But also deviations between these scores and diagnostic quality perceived by radiologists have been observed \cite{NRMetricVsDiagnosticValue}. Blurriness metrics were quite successful in detecting images with reduced quality as perceived by humans in different image domains.

In this paper, we select and present a set of quality metrics (see Tab.\,\ref{tab:nr_metrics}) that could complement reference metrics and detect especially those distortions, which reference metrics can miss. It has often been discussed \cite{contrast_syn_tumor_assess, latent_diffusion_medical}, that error metrics are not sensitive to blurring, which can be problematic because synthesis models may create blurry results. That is why we evaluated a set of blurriness metrics, that do not need a reference. Similar to the learned similarity metrics, also learned quality metrics have shown to provide useful quality scores for natural images \cite{brisque, niqe}. Last, we assessed metrics, that detect MR acquisition artifacts \cite{schuppert2022} or noise \cite{total_variation}.
\subsubsection{Blurriness Metrics}
A large set of metrics has been developed to measure the sharpness or, inversely, the blurriness of images to filter out low-quality images. Methods assessing image blur operate either on the spatial domain, the spectral domain, e.g. through wavelet or fast Fourier transform. In addition, there are learned blur detection methods as well as combinations \cite{review_sharpness}.
Blur assessment methods in the spatial domain, such as the Blur-Effect \cite{blur_effect} or the variance of the Laplacian (VL) \cite{VarLapl},  can exploit local image gradients. Others rely on binary edge detection, with the drawback, that thresholds are needed to decide which pixel belongs to an edge and which one does not. Hence, thresholds need to be adapted for varying intensity value ranges. In general, spectral domain transforms are computationally more costly, so methods on the spatial domain tend to perform faster.

The mean blur (MB) and blur ratio (BR) metrics were jointly \cite{blur2} designed to assess blurriness and edges based on a ratio called inverse blurriness.
A set of blurriness metrics has been derived from the concept of measuring blurred edge widths (BEW) \cite{BlurMarz}.
This idea was extended with a notion of just noticeable blur (JNB) \cite{BlurJNB}, and evaluates the image in smaller blocks.
A further blurriness metric measures the cumulative probability of blur detection (CPBD) \cite{BlurCPBD} extending the approach of JNB.

The incorporated just noticeable blur width is based on experiments with 8-bit integer valued images. Similarly the MB and BR metrics were designed for 8-bit integer valued images. Therefore, our implementation uses a data range parameter for adapting to larger intensity ranges. Further details of the implementations can be found in our published repository at \url{www.github.com/bayer-group/mr-image-metrics}.
\clearpage
\subsubsection{MR Quality Metrics}
In MR images, specific artifacts may appear, which are related to image acquisition and reconstruction. These artifacts may not only appear on real images, but could be reproduced in synthetic images, which is undesirable. Therefore, the use of MR specific quality metrics could efficiently improve validation of MR synthesis models.
In order to select the preferred image from a repeated set of image acquisitions of the same patient, Schuppert et al. \cite{schuppert2022}, evaluated a set of image quality metrics. Mean line correlation (MLC, in \cite{schuppert2022} denoted as "average structural noise") and mean shifted line correlation (MSLC, in \cite{schuppert2022} denoted as "average nyquist ghosting") were revealed to be among the best metrics to predict which image was preferred among repeated acquisitions. Possibly, these metrics are able to detect common MR acquisition artifacts, such as ghosting or motion artifacts, that would lead to repeated acquisitions. 
The MLC metric is defined as the mean correlation between neighboring lines of pixels in an image. 
The MSLC metric is defined as the mean correlation between image lines, that are separated by half of the image width or height respectively.
\subsubsection{Learned Quality Metrics}
Similar to learned reference metrics, also non-reference metrics have been developed from learned image features. The blind/ reference-less image spatial quality evaluator BRISQUE \cite{brisque} leverages 18 spatial image features extracted from distorted training images of the LIVE database \cite{LIVE} annotated with a quality score. A simple support vector machine regression model was trained to predict the annotated quality scores from the extracted set of features. 

The natural image quality evaluator (NIQE)\cite{niqe} does not rely on training with annotated images. Instead, a multi-variate Gaussian model is parameterized from the same set of 18 spatial features, but extracted from two scales. A reference model was parameterized from features from a training set of undistorted images to obtain the multivariate Gaussian model. Images were selected from copyright free Flickr data and from the Berkeley image segmentation database \cite{Berkeley_data}. The NIQE metric assesses the distance of fitted test image parameters to the parameters of the reference model. Due to the characteristics of the training set and assumed differences to MR images regarding intensity value distributions, the NIQE metric may not be directly transferable to MR images. 

\subsubsection{Noise Metrics}
For denoising of images, total variation \cite{total_variation} has been used as a criterion for noisiness. Therefore, mean total variation seems promising as a measure of undesired noise.

Because noise can be reduced by blurring, blurriness metrics might act as inverse noisiness metrics. In other words, an increasing degree of blurriness may correlate with decreasing noise. Inversely, adding noise may disguise blurriness and therefore impair blurriness metrics for image quality assessment.

The non-reference metrics investigated within this study are defined in detail in Supplementary Sec.\,A.3, an overview is shown in Tab.\,\ref{tab:metrics}.
\section{Experiments} 
\label{sec:distortion_experiments}
\begin{figure}[tb]
\centering
\includegraphics{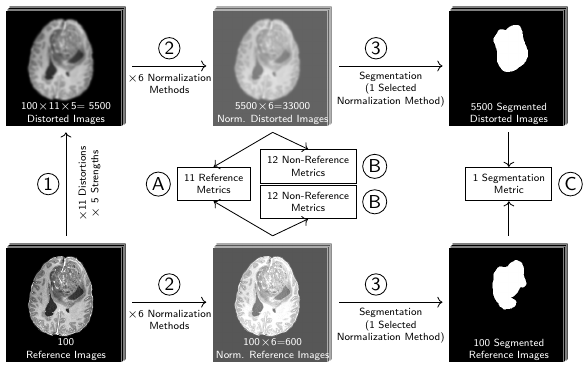}
\caption{Workflow of experiments. (1) 100 reference images were distorted with one of 11 distortions (see Sec.\,\ref{sec:distortions}) with one of five strengths. (2) The distorted images and the reference images were individually normalized with one of six normalization methods (see Sec.\,\ref{sec:normalization_methods}), including no normalization and omitting piece-wise linear (PL) normalization, which depends on a reference dataset. (A) Reference and (B) non-Reference metric scores were obtained from normalized distorted images and normalized reference images. (3) The segmentation model was applied to one normalization method only, because the fully automatic segmentation setup integrated all preprocessing steps, including Zscore normalization.}
\label{fig:workflowexperiments}
\end{figure}
In order to systematically investigate reference and non-reference metrics, we distorted 100 T1-weighted contrast enhanced MR images with 11 different types of distortions in five strengths. For the reference metrics (see Sec.\,\ref{sec:ref_metrics}), the similarity between each distorted image and its undistorted reference was calculated. For the non-reference metrics (see Sec.\,\ref{sec:nr_metrics}), the metric scores for all distorted and undistorted images were assessed. For the segmentation metric (see Sec.\,\ref{sec:seg_metrics}), we trained a model and predicted segmentations for all distorted and undistorted images. The segmentation metric assessed the agreement between segmentations derived from distorted images and segmentations derived from the respective undistorted reference image. In addition, images individually normalized with one of six different normalization methods, including no normalization, leaving the images with raw intensity values. The workflow of the experiments is illustrated in Fig.\,\ref{fig:workflowexperiments}.
In image-to-image tasks, MR source or target images are typically normalized for model training and the synthesized images are generated in this normalized space. Validation of synthesized images can either be performed in this normalized space, such that the normalized target image is used as reference and the synthesized image is assumed to already be normalized appropriately. Another possibility is to invert the previously performed normalization method on the synthesized image to the original intensity range. Then the synthesized image can be compared to the target image in the original intensity range.
In our experiments, we test the metrics in the original and a normalized intensity range. As some distortions slightly or more drastically extend or reduce the intensity range of the reference image, different normalization methods result in different alignment of histograms of the reference and the distorted images.
The LPIPS metric requires an input range of [-1 and 1], therefore we decided to apply Minmax and cMinmax normalization to the required target range of $[-1, 1]$. The DISTS metric requires an input range of $[0, 1]$.. Even though all normalization methods besides Minmax and cMinmax do not satisfy the required input ranges, we did evaluate the metrics after these normalization methods to investigate deviations to the recommended type of normalization.

This experimental setup allows to qualitatively derive the sensitivity of each analyzed metrics to each of the tested distortion types. Assuming that different distortions applied with the same strength should receive the same similarity or quality score, deviations of metric scores between different distortions of the same strengths can be interpreted as strong or weak sensitivity to certain distortions.
\subsection{Data}
We selected the first 100 cases of the Brain Synthesis (BraSyn) 2023 Challenge, available at \url{www.synapse.org/brats2023} \cite{brats1, brats2, brats3}. We further selected only T1-weighted contrast-enhanced (T1c) images. The images all show human brains with glioma tumors. The provided data has already been preprocessed including skull-stripping (removal of the skull), background voxel intensities are set to 0, resampling to a unit mm voxel spacing, registration to a centered brain atlas. For better visualization and reduced computation time, we extracted the centered 2D slice of each 3D volume with a size of 240 $\times$ 240 pixels.
\subsection{Segmentation model for downstream task}
We trained an automatically configuring U-Net based segmentation network \cite{monai_autoseg} on the T1c images of the BraSyn dataset train split. The prediction of three classes (1: whole tumor, 2: tumor core, 3: enhancing tumor) was optimized for 300 epochs, using 1 fold and with a DICE cross entropy loss and deep supervision. The model was selected by the best validation score at epoch 323.
The architecture of the U-Net included five residual blocks, with downsampling factors 1, 2, 2, 4 and 4, initially 32 features and one output channel activated by a sigmoid function per class, resulting in approx. 29 million parameters. The model was trained to segment all three annotated tumor classes.
As a preprocessing step for training and inference, Zscore normalization was applied to the input images. Therefore, no other normalization methods were tested. 
For evaluation of the DSC metric (see Supplementary Sec.\,A.3, Eq.\,(18)), we infered the model on all reference and all distorted images.
\subsection{Distortions}
\label{sec:distortions}
\begin{figure}[thb]
\includegraphics{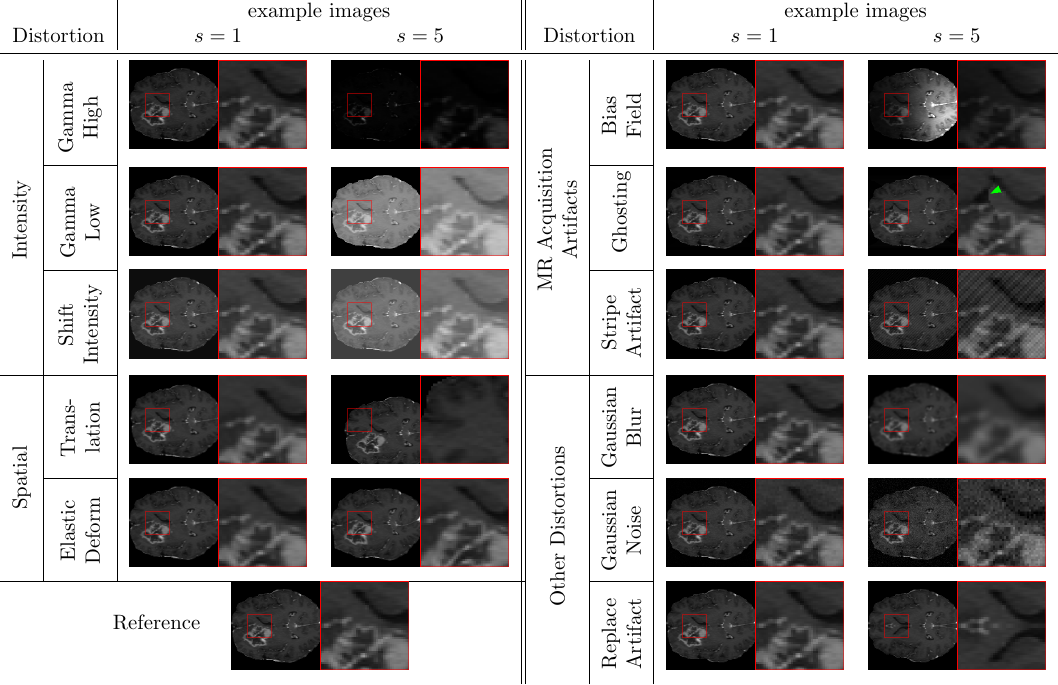}
\caption{Examples of distorted images for lowest strength $s=1$, up to the maximal distortion strength $s=5$. For $s=1$, the distortions are hardly visible and therefore the images appear all the same. All distorted images are displayed with the same intensity range as the reference image, i.e. the range was clipped in case of higher or lower values. The change in the image distorted by ghosting is highlighted by a green arrow. Further examples of distorted images are provided in the Supplementary Figs.\,S.2-S.7}
\label{fig:distortion_examples}
\end{figure}
We selected a wide range of distortions, which we expect to appear with MR image synthesis. The parameters of all distortions were scaled to five increasing strengths, where a strength of one should be a minimal distortion, which is not immediately visible and five a strongly visible distortion, which clearly impedes any diagnosis. We initially scaled the distortion parameters to comparable strengths by a reader study with six experienced researchers (see Supplementary Sec. A5). The final parameters for each distortion are listed in Supplementary Sec.\,A.4. Examples for minimum (strength = 1) and maximum (strength = 5) distortions are shown in Fig.\,\ref{fig:distortion_examples}.

Among the selected distortions, \textbf{Translation} and \textbf{Elastic Deformation} were applied as spatial transforms, that are commonly found, when the reference is not well aligned to the image to be tested. This is frequently the case in image-to-image translation, when the input image was acquired with a different modality or at a different time point. Usually, the patient has moved in between and registration was possibly not sufficient.
Translation was modeled as an equal shift of all pixels along the x and y-axis and parameterized with a fraction of the image width and height. Elastic deformation was modeled by placing a grid with a given number of points on the image, randomly displacing grid points, and linearly interpolating between the new point positions. The displacements were sampled from a normal distribution with increasing parameter $\sigma$ and the number of points was reduced for higher distortion strengths.

Intensity distortions, that shift, stretch or compress the histogram, such as gamma transforms or an intensity shift, can appear between different scanning parameters, because MR does not guarantee a fixed intensity scale. For gamma transforms, images are first normalized by Minmax to range $[0, 1]$. This ensures, that the intensity value range is unchanged under gamma transformation, which is simply potentiating with a parameter $\gamma$. Then, the intensities are scaled back to the original intensity range. We call gamma transforms for $\gamma > 1$ denominated \textbf{Gamma High}, while those with $\gamma<1$ are named \textbf{Gamma Low}. Both types of distortions are parameterized with increasing or decreasing values of $\gamma$ respectively for increasing distortion strengths. \textbf{Intensity shifts} are modeled by adding a fraction of the maximum intensity range to the intensity value of all pixels.

Further distortions, that represent typical acquisition artifacts of MR images are ghosting, stripe and bias field artifacts. \textbf{Ghosting} artifacts appear as shifted copies of the image, arising from erroneous sampling in the frequency space. Scaling a single pixel with an intensity parameter in the frequency space causes artificial \textbf{Stripe Artifacts}. \textbf{Bias fields} appear as low frequency background signals, that we model by multiplying with an exponential of a polynomial function of degree three (see Supplementary Sec.\,A.4 Eq.\,(45)). All of these MR acquisition distortions may moderately expand the intensity range of the distorted image compared to the reference image.

Gaussian noise or Gaussian blurring are not restricted to MR acquisition but, as in most imaging modalities, they are frequently observed and were also analyzed in our study. Gaussian noise adds intensity values randomly sampled from a normal distribution with $\mu=0$ and increasing $\sigma$ to each pixel intensity. Gaussian blur convolves the reference image with a Gaussian filter with increasing $\sigma$.

Last, we investigate the effect of \textbf{Replace Artifacts}, where parts of the image content are replaced, in this case by mirrored regions. In the BraSyn data set, in most cases, there is a tumor in exactly one hemisphere of the brain. By mirroring one hemisphere onto the other one, a second tumor is inserted into, or a tumor is removed from the second hemisphere. We scaled this distortion by mirroring an increasing fraction of the hemisphere. Replacing brain structures in one hemisphere by structure in the other one, simulates the generation of synthetic structures, that were not in the input image. This is a known problem of some synthesis models, e.\,g. of cycleGAN architectures \cite{hallucinate_MR2CT}.
The detection of such synthetically inserted  structures is highly desired for image-to-image translation model validation.
\section{Results}
\label{sec:results}
\subsection{Reference Metrics}
\begin{table}
\caption{Median reference metric values for each distortion evaluated on 100 images and all distortions strength. All reference metric scores shown here were assessed on images without normalization, except LPIPS*, DISTS* and DSC*. The darker the background the higher the sensitivity of the metric to the respective distortion compared to all other distortions. The arrows indicate, if the metric increases ($\uparrow$) or decreases ($\downarrow$) with higher similarity. 
 $^*$: For LPIPS Minmax normalization to [-1, 1] was used, for DISTS Minmax normalization to [0, 1] was applied and DSC was assessed after Zscore normalization and segmentation. Results for all normalization methods can be found in Supplementary Figs.\,S.8-S.15.} 
\label{tab:metrics_median_results}
\begin{tabular}{p{2.4cm}rrrrrrrrrrrr}
metrics \rotatebox[origin=c]{90}{\parbox{0.9cm}{\hspace{0.9cm}}} & \multirow[c]{2}{*}{\metricboxup{SSIM}} & \multirow[c]{2}{*}{\metricboxup{MS-SSIM}} & \multirow[c]{2}{*}{\metricboxup{CW-SSIM}} & \multirow[c]{2}{*}{\metricboxup{PSNR}} & \multirow[c]{2}{*}{\metricboxdown{MAE}} & \multirow[c]{2}{*}{\metricboxdown{MSE}} & \multirow[c]{2}{*}{\metricboxdown{NMSE}} & \multirow[c]{2}{*}{\metricboxdown{LPIPS$^*$}} & \multirow[c]{2}{*}{\metricboxdown{DISTS$^*$}} & \multirow[c]{2}{*}{\metricboxup{NMI}} & \multirow[c]{2}{*}{\metricboxup{PCC}} & \multirow[c]{2}{*}{\metricboxup{DSC$^*$}} \\
Distortions &  &  &  &  &  &  &  &  &  &  &  &  \\
Bias Field & \cellcolor{gray!2}\color{black}0.96 & \cellcolor{gray!21}\color{black}0.89 & \cellcolor{gray!24}\color{black}0.92 & \cellcolor{gray!57}\color{white}28.05 & \cellcolor{gray!15}\color{black}204.43 & \cellcolor{gray!14}\color{black} 2.00$\cdot 10^{5}$ & \cellcolor{gray!26}\color{black}324.16 & \cellcolor{gray!13}\color{black}0.06 & \cellcolor{gray!25}\color{black}0.08 & \cellcolor{gray!78}\color{white}1.32 & \cellcolor{gray!31}\color{black}0.91 & \cellcolor{gray!11}\color{black}0.92 \\
\cline{1-13}
Ghosting & \cellcolor{gray!7}\color{black}0.93 & \cellcolor{gray!1}\color{black}0.99 & \cellcolor{gray!6}\color{black}0.98 & \cellcolor{gray!0}\color{black}42.23 & \cellcolor{gray!1}\color{black}34.58 & \cellcolor{gray!0}\color{black} 3.58$\cdot 10^{3}$ & \cellcolor{gray!0}\color{black}3.52 & \cellcolor{gray!4}\color{black}0.02 & \cellcolor{gray!35}\color{black}0.11 & \cellcolor{gray!61}\color{white}1.46 & \cellcolor{gray!0}\color{black}1.00 & \cellcolor{gray!3}\color{black}0.97 \\
\cline{1-13}
Stripe Artifact & \cellcolor{gray!67}\color{white}0.52 & \cellcolor{gray!2}\color{black}0.98 & \cellcolor{gray!1}\color{black}0.99 & \cellcolor{gray!34}\color{black}33.72 & \cellcolor{gray!6}\color{black}94.77 & \cellcolor{gray!1}\color{black} 2.07$\cdot 10^{4}$ & \cellcolor{gray!1}\color{black}22.17 & \cellcolor{gray!25}\color{black}0.11 & \cellcolor{gray!100}\color{white}0.33 & \cellcolor{gray!71}\color{white}1.37 & \cellcolor{gray!4}\color{black}0.99 & \cellcolor{gray!14}\color{black}0.90 \\
\cline{1-13}
Gaussian Blur & \cellcolor{gray!0}\color{black}0.98 & \cellcolor{gray!0}\color{black}1.00 & \cellcolor{gray!0}\color{black}1.00 & \cellcolor{gray!16}\color{black}38.27 & \cellcolor{gray!0}\color{black}21.11 & \cellcolor{gray!0}\color{black} 4.65$\cdot 10^{3}$ & \cellcolor{gray!0}\color{black}5.17 & \cellcolor{gray!16}\color{black}0.07 & \cellcolor{gray!36}\color{black}0.12 & \cellcolor{gray!54}\color{white}1.52 & \cellcolor{gray!1}\color{black}1.00 & \cellcolor{gray!10}\color{black}0.93 \\
\cline{1-13}
Gaussian Noise & \cellcolor{gray!57}\color{white}0.58 & \cellcolor{gray!15}\color{black}0.92 & \cellcolor{gray!6}\color{black}0.98 & \cellcolor{gray!43}\color{black}31.54 & \cellcolor{gray!13}\color{black}181.33 & \cellcolor{gray!3}\color{black} 5.16$\cdot 10^{4}$ & \cellcolor{gray!3}\color{black}47.68 & \cellcolor{gray!100}\color{white}0.44 & \cellcolor{gray!99}\color{white}0.32 & \cellcolor{gray!90}\color{white}1.21 & \cellcolor{gray!10}\color{black}0.97 & \cellcolor{gray!20}\color{black}0.86 \\
\cline{1-13}
Replace Artifact & \cellcolor{gray!5}\color{black}0.95 & \cellcolor{gray!10}\color{black}0.94 & \cellcolor{gray!22}\color{black}0.92 & \cellcolor{gray!46}\color{black}30.80 & \cellcolor{gray!1}\color{black}35.35 & \cellcolor{gray!3}\color{black} 5.10$\cdot 10^{4}$ & \cellcolor{gray!3}\color{black}47.77 & \cellcolor{gray!8}\color{black}0.04 & \cellcolor{gray!16}\color{black}0.05 & \cellcolor{gray!41}\color{black}1.64 & \cellcolor{gray!8}\color{black}0.98 & \cellcolor{gray!40}\color{black}0.71 \\
\cline{1-13}
Gamma High & \cellcolor{gray!13}\color{black}0.89 & \cellcolor{gray!32}\color{black}0.84 & \cellcolor{gray!58}\color{white}0.80 & \cellcolor{gray!81}\color{white}22.19 & \cellcolor{gray!25}\color{black}315.44 & \cellcolor{gray!23}\color{black} 3.31$\cdot 10^{5}$ & \cellcolor{gray!24}\color{black}299.66 & \cellcolor{gray!18}\color{black}0.08 & \cellcolor{gray!41}\color{black}0.14 & \cellcolor{gray!23}\color{black}1.79 & \cellcolor{gray!8}\color{black}0.98 & \cellcolor{gray!14}\color{black}0.90 \\
\cline{1-13}
Gamma Low & \cellcolor{gray!3}\color{black}0.96 & \cellcolor{gray!11}\color{black}0.94 & \cellcolor{gray!23}\color{black}0.92 & \cellcolor{gray!86}\color{white}21.07 & \cellcolor{gray!25}\color{black}312.06 & \cellcolor{gray!23}\color{black} 3.26$\cdot 10^{5}$ & \cellcolor{gray!22}\color{black}273.75 & \cellcolor{gray!8}\color{black}0.04 & \cellcolor{gray!26}\color{black}0.09 & \cellcolor{gray!21}\color{black}1.81 & \cellcolor{gray!2}\color{black}0.99 & \cellcolor{gray!15}\color{black}0.89 \\
\cline{1-13}
Shift Intensity & \cellcolor{gray!100}\color{white}0.29 & \cellcolor{gray!2}\color{black}0.99 & \cellcolor{gray!26}\color{black}0.91 & \cellcolor{gray!100}\color{white}17.69 & \cellcolor{gray!100}\color{white} 1.17$\cdot 10^{3}$ & \cellcolor{gray!100}\color{white} 1.37$\cdot 10^{6}$ & \cellcolor{gray!100}\color{white} 1.20$\cdot 10^{3}$ & \cellcolor{gray!0}\color{black}0.00 & \cellcolor{gray!0}\color{black}-0.00 & \cellcolor{gray!0}\color{black}2.00 & \cellcolor{gray!0}\color{black}1.00 & \cellcolor{gray!0}\color{black}1.00 \\
\cline{1-13}
Translation & \cellcolor{gray!37}\color{black}0.72 & \cellcolor{gray!100}\color{white}0.52 & \cellcolor{gray!100}\color{white}0.65 & \cellcolor{gray!85}\color{white}21.24 & \cellcolor{gray!24}\color{black}306.77 & \cellcolor{gray!40}\color{black} 5.61$\cdot 10^{5}$ & \cellcolor{gray!42}\color{black}509.30 & \cellcolor{gray!54}\color{white}0.24 & \cellcolor{gray!25}\color{black}0.08 & \cellcolor{gray!100}\color{white}1.13 & \cellcolor{gray!100}\color{white}0.72 & \cellcolor{gray!100}\color{white}0.29 \\
\cline{1-13}
Elastic Deform & \cellcolor{gray!8}\color{black}0.92 & \cellcolor{gray!7}\color{black}0.96 & \cellcolor{gray!11}\color{black}0.96 & \cellcolor{gray!46}\color{black}30.91 & \cellcolor{gray!3}\color{black}56.64 & \cellcolor{gray!2}\color{black} 4.42$\cdot 10^{4}$ & \cellcolor{gray!3}\color{black}41.92 & \cellcolor{gray!9}\color{black}0.04 & \cellcolor{gray!25}\color{black}0.08 & \cellcolor{gray!69}\color{white}1.39 & \cellcolor{gray!9}\color{black}0.97 & \cellcolor{gray!11}\color{black}0.92 \\
\cline{1-13}
\end{tabular}
\end{table}
\begin{table}
\caption{Comparison of \textbf{relative} metric scores for selected distortions for comparison of normalization methods. As a relative metric score, the median  of one selected distortion is divided by the median metric score of all distortions. The gray background indicates higher sensitivity in  the comparison of two normalization methods. LPIPS, DISTS and DSC metric are not shown, due to fixed recommendations regarding the normalization method. Normalization methods and distortions were selected, where the similarity uniformly changed for between two normalization methods within one type of distortion and for all reference metrics. The arrows indicate, if the metrics metric increases ($\uparrow$) or decreases ($\downarrow$) with increasing similarity.}
\label{tab:metrics_norm_results}
\begin{tabular}{p{2.4cm}crrrrrrrrrrrr}
 \hspace{0.5cm} metrics \rotatebox[origin=c]{90}{\parbox{1.2cm}{\hspace{1.2cm}}} &&\multirow[c]{2}{*}{\metricboxup{SSIM}}&\multirow[c]{2}{*}{\metricboxup{MS-SSIM}}&\multirow[c]{2}{*}{\metricboxup{CW-SSIM}}&\multirow[c]{2}{*}{\metricboxup{PSNR}}&\multirow[c]{2}{*}{\metricboxdown{MSE}}&\multirow[c]{2}{*}{\metricboxdown{NMSE}}&\multirow[c]{2}{*}{\metricboxdown{MAE}}&\multirow[c]{2}{*}{\metricboxup{NMI}}&\multirow[c]{2}{*}{\metricboxup{PCC}}\\
Distortions & Norm. &&&&&&&&\\
\hline
\multirow[l]{2}{*}{Gamma High}&w/o Norm.&\cellcolor{gray!50}0.98&\cellcolor{gray!50}0.87&\cellcolor{gray!50}0.83&\cellcolor{gray!50}0.75&\cellcolor{gray!50}3.08&\cellcolor{gray!50}5.77&\cellcolor{gray!50}4.52&\cellcolor{gray!50}1.22&\cellcolor{gray!50}0.99\\
 &$\mathrm{cMinmax}_{5\%}$&1.07&1.02&0.99&1.00&1.17&0.98&0.96&1.31&1.00\\
\hline
\multirow[l]{2}{*}{Gamma Low}&w/o Norm.&\cellcolor{gray!50}1.05&\cellcolor{gray!50}0.97&\cellcolor{gray!50}0.95&\cellcolor{gray!50}0.71&\cellcolor{gray!50}3.05&\cellcolor{gray!50}5.69&\cellcolor{gray!50}4.13&\cellcolor{gray!50}1.23&1.01\\
&$\mathrm{cMinmax}_{5\%}$&1.07&1.02&1.00&1.13&0.77&0.48&0.47&1.31&1.01\\
\hline
\multirow[l]{2}{*}{Gaussian Noise}&w/o Norm.&0.64&0.96&1.02&1.06&1.77&0.90&0.72&0.82&0.98\\
&Minmax&\cellcolor{gray!50}0.26&\cellcolor{gray!50}0.95&\cellcolor{gray!50}0.95&\cellcolor{gray!50}0.71&\cellcolor{gray!50}4.81&\cellcolor{gray!50}6.38&\cellcolor{gray!50}6.90&0.82&0.98\\
\hline
\multirow[l]{2}{*}{Ghosting}&w/o Norm.&1.03&1.03&1.02&1.42&0.34&0.06&0.05&0.99&1.01\\
&Minmax&\cellcolor{gray!50}0.39&\cellcolor{gray!50}1.02&\cellcolor{gray!50}1.00&\cellcolor{gray!50}0.95&\cellcolor{gray!50}2.32&\cellcolor{gray!50}1.37&\cellcolor{gray!50}1.42&0.99&1.01\\
\hline
\multirow[l]{2}{*}{Stripe Artifact}&w/o Norm.&\cellcolor{gray!50}0.57&1.02&1.03&1.14&0.93&0.36&0.33&0.93&1.00\\
&ZScore&0.84&\cellcolor{gray!50}1.01&\cellcolor{gray!50}1.02&\cellcolor{gray!50}0.98&\cellcolor{gray!50}1.78&\cellcolor{gray!50}0.95&\cellcolor{gray!50}0.95&0.93&1.00\\
\end{tabular}
\end{table}
The complete results for all reference metrics, all distortion strengths and all normalization methods are contained in the Supplementary Figs.\,S.8-S.10. There, for each metric and normalization method, the trends of all distortions for increasing strengths are shown. The selected and compressed results in Tab.\,\ref{tab:metrics_median_results}, contain the median metric values over all strengths and images for one selected and recommended normalization method. For LPIPS and DISTS, images we show results for Minmax normalization to range $[-1,1]$ and $[0,1]$ respectively as recommended by the authors of the used implementations. For DSC, Zscore normalization was part of the segmentation process. For all other metrics, we show results without normalization as these seem representative and we were not aware of any recommended normalization method.
Tab.\,\ref{tab:metrics_median_results} only contains the DSC on the foreground (union of all three tumor classes), as the DSC segmentation scores for all three classes are very similar (see Supplementary Fig.\,S.15).
In the performed experiments, dependencies between distortions of different strengths and metric scores can be observed. The majority of metrics indicate decreasing similarity or decreasing quality for increasing distortion strengths. This can be described as sensitivity of a metric to a distortion or an effect of a distortion to a metric. Specifically, stronger changes of metrics scores observed for one distortion, compared to other distortions can be interpreted as a higher sensitivity of the metric or a stronger effect of the distortion to this metric.
The results are described in the following, ordered by distortions.

\textbf{Bias Field} has a moderate effect on most metrics. MS-SSIM is much more sensitive to simulated bias field artifacts than simple SSIM or CW-SSIM. The error metrics show clearly increased (dissimilarity) scores, while the LPIPS score is hardly effected. Compared to other distortions, PCC drops noticeably with bias field distortions.
\textbf{Ghosting} generally has a weak effect on most metrics, except on NMI and DISTS.
\textbf{Stripe Artifacts} strongly influence a subset of metrics, including SSIM, LPIPS, DISTS and NMI, while most other metrics are not sensitive to this type of distortion.
\textbf{Blurring} is hardly accounted for by most metrics, the strongest changes can be observed by the DISTS and NMI metrics.
Regarding \textbf{Gaussian Noise}, SSIM is very sensitive, while MS-SSIM and CW-SSIM are not. The effect on error metrics is limited, while both learned metrics and NMI clearly indicate dissimilarity.
\textbf{Replace Artifacts} are hardly detected by most metrics. Besides DSC, which aligns well with the distortion strength, PSNR and NMI are most sensitive. 
\textbf{Gamma transforms} with an increasing $\gamma > 1$ (Gamma High) or a decreasing $\gamma < 1$ (Gamma Low) similarly influence all metrics. All of them, except PSNR, assess the distortions with $\gamma > 1$ as stronger. 
Regarding constant \textbf{Intensity Shifts}, NMI and PCC are invariant. Their scores reflect perfect similarity. The same holds, when the images are normalized by any of the five normalization methods (see Supplementary Figs. S.8-S.15). Particularly for LPIPS and DISTS, Minmax normalization is recommended as standard. For all other metrics, except MS-SSIM and CW-SSIM, intensity shifts substantially decrease similarity.
\textbf{Translation} strongly reduces the assessed similarity for all metrics besides DISTS. Even very small translations of strength $s=1$, which corresponds to a 1\% shift (2.4 pixels in our experiments), are clearly noticeable, as shown in Supplementary Sec.\,B.1. Only CW-SSIM is quite insensitive with respect to small translations, but is still very sensitive, when translation is strong.
Compared to translation, \textbf{Elastic Deforms} only influence similarity metrics decently. Even stronger deformations have less impact on the metric scores, than the weakest translation does.

There are some coherent observations regarding the \textbf{normalization} method and certain types of distortions. Selected results are shown in Tab.\,\ref{tab:metrics_norm_results}.
$\mathrm{cMinmax}_{5\%}$ normalization reduces gamma transforms, such that most metrics are less sensitive to gamma transforms after $\mathrm{cMinmax}_{5\%}$ normalization, compared to their standard normalization method.
Minmax normalization amplifies Gaussian noise,  Zscore normalization amplifies stripe artifacts.

Gaussian noise and stripe artifacts most strongly decrease the DSC. Translated segmentations have decreasing overlap and thereby very low DSC. Tumors, which were inserted or removed by the Replace Reflect distortion are well indicated by decreasing DSC.
\subsection{Non-Reference Metrics}
\begin{table}
\caption{Median non-reference metric values for each distortion over 100 images. All non-reference metrics were assessed with binning normalization, which best resembles the intended use on 8-bit integer valued images for which most of these metrics were developed. Results for all normalization methods can be found in Supplementary Figs.\,S.11-S.14. The darker the background the higher the difference to the median metric value of the reference images. The arrows indicate, if the metric increases ($\uparrow$) or decreases ($\downarrow$ with image quality.}
\label{tab:nr_metrics_median_results}
\begin{tabular}{p{2.4cm}rrrrrrrrrrrr}
metrics \rotatebox[origin=c]{90}{\parbox{1.2cm}{\hspace{1.2cm}}} & \multirow[c]{2}{*}{\metricboxup{BE}} & \multirow[c]{2}{*}{\metricboxup{BR}} & \multirow[c]{2}{*}{\metricboxup{MB}} & \multirow[c]{2}{*}{\metricboxdown{VL}} & \multirow[c]{2}{*}{\metricboxdown{MTV}} & \multirow[c]{2}{*}{\metricboxup{BEW}} & \multirow[c]{2}{*}{\metricboxup{JNB}} & \multirow[c]{2}{*}{\metricboxdown{CPBD}} & \multirow[c]{2}{*}{\metricboxdown{BRISQUE}} & \multirow[c]{2}{*}{\metricboxdown{NIQE}} & \multirow[c]{2}{*}{\metricboxdown{MSLC}} & \multirow[c]{2}{*}{\metricboxdown{MLC}} \\
Distortions &  &  &  &  &  &  &  &  &  &  &  &  \\
Bias Field & \cellcolor{gray!7}\color{black}0.34 & \cellcolor{gray!4}\color{black}2.20 & \cellcolor{gray!0}\color{black}4.55 & \cellcolor{gray!38}\color{black} 5.71$\cdot 10^{5}$ & \cellcolor{gray!31}\color{black}127.29 & \cellcolor{gray!5}\color{black}3.26 & \cellcolor{gray!10}\color{black}15.96 & \cellcolor{gray!2}\color{black}0.61 & \cellcolor{gray!5}\color{black}69.46 & \cellcolor{gray!0}\color{black}23.10 & \cellcolor{gray!3}\color{black}0.16 & \cellcolor{gray!0}\color{black}0.97 \\
\cline{1-13}
Ghosting & \cellcolor{gray!2}\color{black}0.33 & \cellcolor{gray!68}\color{white}5.89 & \cellcolor{gray!0}\color{black}0.93 & \cellcolor{gray!19}\color{black} 2.95$\cdot 10^{5}$ & \cellcolor{gray!25}\color{black}104.25 & \cellcolor{gray!3}\color{black}3.50 & \cellcolor{gray!3}\color{black}24.89 & \cellcolor{gray!1}\color{black}0.58 & \cellcolor{gray!8}\color{black}53.81 & \cellcolor{gray!1}\color{black}22.56 & \cellcolor{gray!19}\color{black}0.19 & \cellcolor{gray!20}\color{black}0.84 \\
\cline{1-13}
Stripe Artifact & \cellcolor{gray!45}\color{black}0.24 & \cellcolor{gray!31}\color{black}0.07 & \cellcolor{gray!99}\color{white} 3.59$\cdot 10^{3}$ & \cellcolor{gray!80}\color{white} 1.20$\cdot 10^{6}$ & \cellcolor{gray!81}\color{white}327.94 & \cellcolor{gray!27}\color{black}2.66 & \cellcolor{gray!25}\color{black}6.72 & \cellcolor{gray!49}\color{black}0.89 & \cellcolor{gray!82}\color{white}155.84 & \cellcolor{gray!97}\color{white}93.27 & \cellcolor{gray!83}\color{white}0.28 & \cellcolor{gray!97}\color{white}0.38 \\
\cline{1-13}
Gaussian Blur & \cellcolor{gray!39}\color{black}0.41 & \cellcolor{gray!7}\color{black}2.37 & \cellcolor{gray!0}\color{black}5.39 & \cellcolor{gray!1}\color{black} 2.75$\cdot 10^{4}$ & \cellcolor{gray!16}\color{black}68.80 & \cellcolor{gray!72}\color{white}5.42 & \cellcolor{gray!74}\color{white}70.00 & \cellcolor{gray!50}\color{white}0.28 & \cellcolor{gray!9}\color{black}73.69 & \cellcolor{gray!1}\color{black}24.97 & \cellcolor{gray!17}\color{black}0.19 & \cellcolor{gray!2}\color{black}0.98 \\
\cline{1-13}
Gaussian Noise & \cellcolor{gray!60}\color{white}0.21 & \cellcolor{gray!26}\color{black}0.40 & \cellcolor{gray!0}\color{black}5.92 & \cellcolor{gray!101}\color{white} 1.51$\cdot 10^{6}$ & \cellcolor{gray!113}\color{white}459.26 & \cellcolor{gray!10}\color{black}3.13 & \cellcolor{gray!17}\color{black}11.57 & \cellcolor{gray!26}\color{black}0.75 & \cellcolor{gray!17}\color{black}43.70 & \cellcolor{gray!18}\color{black}37.06 & \cellcolor{gray!10}\color{black}0.15 & \cellcolor{gray!66}\color{white}0.56 \\
\cline{1-13}
Replace Artifact & \cellcolor{gray!0}\color{black}0.33 & \cellcolor{gray!0}\color{black}1.92 & \cellcolor{gray!0}\color{black}5.44 & \cellcolor{gray!22}\color{black} 3.33$\cdot 10^{5}$ & \cellcolor{gray!25}\color{black}104.79 & \cellcolor{gray!1}\color{black}3.36 & \cellcolor{gray!0}\color{black}22.51 & \cellcolor{gray!0}\color{black}0.59 & \cellcolor{gray!0}\color{black}63.42 & \cellcolor{gray!0}\color{black}23.20 & \cellcolor{gray!4}\color{black}0.17 & \cellcolor{gray!0}\color{black}0.97 \\
\cline{1-13}
Gamma High & \cellcolor{gray!15}\color{black}0.30 & \cellcolor{gray!2}\color{black}1.76 & \cellcolor{gray!0}\color{black}5.13 & \cellcolor{gray!10}\color{black} 1.51$\cdot 10^{5}$ & \cellcolor{gray!13}\color{black}57.30 & \cellcolor{gray!6}\color{black}3.24 & \cellcolor{gray!14}\color{black}13.80 & \cellcolor{gray!4}\color{black}0.62 & \cellcolor{gray!5}\color{black}69.61 & \cellcolor{gray!0}\color{black}23.59 & \cellcolor{gray!16}\color{black}0.15 & \cellcolor{gray!2}\color{black}0.95 \\
\cline{1-13}
Gamma Low & \cellcolor{gray!7}\color{black}0.34 & \cellcolor{gray!5}\color{black}1.59 & \cellcolor{gray!3}\color{black}114.79 & \cellcolor{gray!37}\color{black} 5.62$\cdot 10^{5}$ & \cellcolor{gray!31}\color{black}127.82 & \cellcolor{gray!1}\color{black}3.46 & \cellcolor{gray!4}\color{black}25.43 & \cellcolor{gray!2}\color{black}0.57 & \cellcolor{gray!0}\color{black}62.69 & \cellcolor{gray!0}\color{black}22.86 & \cellcolor{gray!5}\color{black}0.18 & \cellcolor{gray!1}\color{black}0.97 \\
\cline{1-13}
Shift Intensity & \cellcolor{gray!0}\color{black}0.33 & \cellcolor{gray!0}\color{black}1.93 & \cellcolor{gray!0}\color{black}5.32 & \cellcolor{gray!23}\color{black} 3.44$\cdot 10^{5}$ & \cellcolor{gray!25}\color{black}105.77 & \cellcolor{gray!0}\color{black}3.41 & \cellcolor{gray!0}\color{black}22.75 & \cellcolor{gray!0}\color{black}0.59 & \cellcolor{gray!0}\color{black}62.91 & \cellcolor{gray!0}\color{black}23.56 & \cellcolor{gray!0}\color{black}0.17 & \cellcolor{gray!0}\color{black}0.97 \\
\cline{1-13}
Translation & \cellcolor{gray!14}\color{black}0.36 & \cellcolor{gray!3}\color{black}2.14 & \cellcolor{gray!0}\color{black}4.62 & \cellcolor{gray!8}\color{black} 1.27$\cdot 10^{5}$ & \cellcolor{gray!21}\color{black}90.11 & \cellcolor{gray!27}\color{black}4.17 & \cellcolor{gray!20}\color{black}35.52 & \cellcolor{gray!17}\color{black}0.48 & \cellcolor{gray!0}\color{black}63.72 & \cellcolor{gray!2}\color{black}21.91 & \cellcolor{gray!4}\color{black}0.17 & \cellcolor{gray!1}\color{black}0.98 \\
\cline{1-13}
Elastic Deform & \cellcolor{gray!13}\color{black}0.36 & \cellcolor{gray!3}\color{black}2.13 & \cellcolor{gray!0}\color{black}4.64 & \cellcolor{gray!10}\color{black} 1.55$\cdot 10^{5}$ & \cellcolor{gray!22}\color{black}90.27 & \cellcolor{gray!30}\color{black}4.24 & \cellcolor{gray!21}\color{black}36.28 & \cellcolor{gray!18}\color{black}0.48 & \cellcolor{gray!0}\color{black}62.99 & \cellcolor{gray!0}\color{black}23.05 & \cellcolor{gray!8}\color{black}0.18 & \cellcolor{gray!1}\color{black}0.97 \\
\cline{1-13}
\end{tabular}
\end{table}
The results for all non-reference metrics, all distortion strengths and all normalization methods are given in Supplementary Figs.\,S.11-S.14. The median metric scores are summarized in Tab.\,\ref{tab:nr_metrics_median_results} for images, that were normalized by Binning, as most of the metrics were designed for 8-bit images.
This assures, that the metrics are most likely used as intended and sensitivity to a type of distortion is expected to be more consistent. Large differences between normalization methods are found for MB, BR and NIQE.
We describe the results ordered by quality metric groups.

All \textbf{Blurriness and Noisiness Metrics}, except MB, can distinguish well between different strengths of blurring. At the same time, the blurriness scores diverge in the opposite direction for increasing strengths of stripe artifacts and Gaussian noise. Only BEW scores are not influenced much by these distortions.
All blurriness and noisiness metrics show coherent deviations for images with spatial Translation (see also Supplementary Figs.\,S.12-S.13).
\textbf{MR Quality Metrics}, i.e. MLC and MSLC, clearly identify stripe artifacts. MLC is very sensitive to noise, but generally hardly reflects the distortion level. MSLC can identify ghosting, best with $\mathrm{cMinmax}_{5\%}$ normalization (see also Supplementary Fig.\,S.13).
The \textbf{Learned Quality Metrics} BRISQUE and NIQE clearly attest low quality to images with stripe artifacts. However, the BRISQUE score indicates higher quality for very weak stripe artifacts and for images with Gaussian noise or ghosting (see also Supplementary Fig.\,S.14). The NIQE score indicates lower quality for Gaussian noise.

\section{Discussion}
\label{sec:discussion}
The experiments demonstrate specific weaknesses of the popular SSIM and PSNR metrics. They are strongly decreased by constant intensity shifts if no normalization is applied. PSNR is very dependent on the kind of normalization, which complicates its use as a comprehensive metric for comparing studies of different authors. This has been pointed out before\cite{MR_normalization}, but the normalization parameters are still not reported in many papers about medical synthesis models \cite{SSIM_properUse}. SSIM and PSNR underestimate blurring and thereby favor blurred images over differently distorted images. This is inline with reported findings of previous studies \cite{DyeFreeNet2020, MSE_loveit_or_leaveit}.
SSIM and PSNR are both very sensitive to spatial translation, which is a frequently occurring issue in paired image-to-image translation. The source image and the reference image may not acquired by the same hardware or not immediately at the same time points and therefore the patient may not have the same location in both images.
The proper use of the data range parameter of SSIM for medical images and potential biases have been investigated \cite{SSIM_properUse}, suggesting to use the dataset minimum and maximum values. In existing implementations of SSIM or PSNR, the default data range parameter might not be appropriate. 
For SSIM, although commonly applied on Zscore normalized MR images, a bias was demonstrated\cite{SSIM_properUse} with negative values, which we did not specifically investigate.
Even though the generation of synthetic structures is an issue in image-to-image translation \cite{hallucinate_MR2CT}, highly relevant replace artifacts are not sufficiently assessed by SSIM and PSNR.
Unexpected behaviour of SSIM and PSNR as mentioned above was illustrated by examples \cite{ref_metric_pitfalls}.

So how can these weaknesses be overcome?
 CW-SSIM is able to ignore small translations, due to calculation in the complex-wavelet domain. The DISTS metric also successfully focuses more on texture than spatial alignment. 
Precise registration would also strongly improve similarity assessment. However, interpolation may introduce blurring as shown by coherent variations of assessed quality by all non-reference quality metrics on spatially translated images. Our experiments also suggest, that a few strong local elastic deformations have a lot less impact on similarity metrics than rigid translations. At the same time, rigid registration is easier to solve via optimization than elastic registration, because only a few parameters need to be determined.

Replace artifacts remain underestimated by most similarity metrics. For those artifacts, that resemble structures of diagnostic interest, the evaluation of segmentations with a specific segmentation model is useful. In our evaluation, we use a segmentation model, that was trained to detect different tumor regions. It successfully detects replace artifacts, where the tumor is doubled or removed. In this study, we did not perform further stability tests of the segmentation model. The DSC scores evaluated on the segmented images generally very well represent the expected differences between the reference and its distorted versions and manual inspection of selected segmentations did not uncover any obvious segmentation failures.

The non-reference quality metrics can give valuable additional information about the quality of synthesized images. Blurring is easily and reliably detected by all blurriness metrics. For most metrics, assessed blurriness also decreases with other distortions, such as Gaussian noise, ghosting or gamma transforms, when these increase image contrast.

For assessing ghosting, the MR quality metrics MLC and MSLC were evaluated.
The MLC metric represents the line-wise correlation between neighboring lines. The reference value in our experiments lies at 96\%, which is probably caused by high anatomical consistency in the present pixel spacing. It strongly decreases for stripe artifacts, which fits to the fact, that the stripes change relations between local image intensities and are not oriented along the x- or y-axis. Random Gaussian noise reduces statistical correlation and thereby also MLC. Ghosting also reduces MLC, as it additionally distorts image intensities locally.
The MSLC metric only slightly increases with ghosting, a bit more using cMinmax normalization. One reason could be the relative weak scaling of the ghost intensity in our experiments. In contrast, stripe artifacts significantly increased the MSLC metrics. Compared to MLC, by coincidence, stripes seem to be in the same phase at the half-image width distance and thereby drastically increase line-wise correlation.

Although the result tables for reference metrics (Tab.\,\ref{tab:metrics_median_results}) and for non-reference metrics (Tab.\,\ref{tab:nr_metrics_median_results}) show absolute metric scores with and without normalization, both are representative for all results. In our experiments, reference images and distorted images had similar intensity ranges. Of course, MSE, MAE and NMSE yield much larger values without normalization, but by filling the background of the result tables with different shades of gray we emphasize the relative sensitivity of the metrics, and we found that the relative sensitivity is very similar between all normalizations for all metrics, with three exceptions: (1) Intensity shift is fully removed by any normalization leading to perfect similarity or equal quality to the reference. (2) BR, MB and NIQE display large differences between normalization methods and do not seem to measure blurriness or quality as intended when applied on non-8-bit integer valued images, which could be a problem of the adapted implementation, but also a problem of metric design. For all other non-reference metrics, the relative sensitivity does not seem to change with normalization. (3) We highlighted all uniform effects of normalization methods to reference metrics in Tab.\,\ref{tab:metrics_norm_results}, where we observe that certain normalization methods amplify or mitigate certain distortions. Gaussian noise and Ghosting, which extend the intensity value range compared to the reference image, are amplified by Minmax normalization, because the extended intensity value range is more strongly compressed. Stripe artifacts shift image mean values and therefore Zscore normalization creates shifts between originally corresponding intensity values. Gamma transforms, that compress intensity values at very high and low values are mitigated by cMinMax normalization, as this stretches intensities apart again. Therefore, normalization methods can be used to improve similarity or quality of images. Further distortions could be reduced by other methods, e.g.
Bias field correction \cite{biasfieldcorr}, which was designed to remove bias field artifacts.

Our experiments were performed on T1-weighted contrast-enhanced MR images of the human brain and are therefore restricted to the BraSyn dataset. Future work should include a much broader set of MR images with other sequences and body regions to make sure, that the results are valid more generally.
However, we assume, that the results are transferable to most MR sequences, because the characteristics, that are critical for metric assessment, are equally present (large value range, non-quantitative measurement, types of artifacts).
The background in the BraSyn dataset takes up a large fraction of the image and was specifically preprocessed and set to 0. These factors largely influence some of the metrics, especially SSIM, PSNR and the error based metrics and therefore other data could yield different experimental results. We expect that using masks with these metrics will change absolute metric scores, but relative observations probably persist. However, the application of masks with metrics, that do not assess similarity or quality on a per-pixel basis, but require a neighborhood, is still to be implemented as future work. For the segmentation related DSC, taking different sub-regions into account, is already being explored \cite{multiregionDSC}.
Furthermore, we did not investigate differences between a slice-wise 2D and a 3D application of the selected metrics. This should be considered specifically for MR images, and has already been reported to have significant impact \cite{SSIM_properUse}.

The distortions analyzed in this study are to different degrees realistic. Ghosting \cite{ghosting}, Stripe artifacts \cite{stripeArtifact} and Bias field \cite{bias_field} were performed by simulation methods based on MR physics. Other distortions are perhaps less realistic, but account for important problems when acquiring and processing MR images. While other studies \cite{kastryulin, ding2020optim} collected real image outputs of image synthesis models, we applied well-defined distortions in an isolated manner. This allows to better distinguish sensitivity of metrics between different distortions and to better understand specific metric properties.

In comparison to other metric benchmarks\cite{LIVE, tid_dataset}, we do not compare the metric scores to human quality assessment. For each type of distortion, we tried to select five comparable strengths. As shown in Fig.\,\ref{fig:distortion_examples}, the lowest and highest strengths were scaled to appear almost not visible or to decrease quality to an equally poor level. Even though distortion parameters might not be scaled perfectly to human perception, we assume, that the overall qualitative observations about which metrics are most sensitive towards which kind of distortions and the following conclusions are still valid.

We are aware, that absolute values of different metrics cannot be directly compared, because most metrics relate non-linearly to human perception \cite{tid_dataset, LIVE}. Therefore, rank correlation is often utilized to compare different metrics. However, we provide absolute scores, which better display how far apart they are for single distortions. At the same time, we consider the ranking of different distortion types for each metric to derive the sensitivity. For the observation of different dynamics with distortion strengths we refer the reader to the Supplementary Figs.\,S.8-S.15.

After LPIPS, a huge amount of deep-learning based reference and non-reference methods, among them PieApp \cite{PieApp}, MetaIQA \cite{MetaIQA}, P2P-BM \cite{P2P-BM}, HyperIQA \cite{HyperIQA} and AHIQ \cite{AHIQ}, has been proposed in the recent years, which were able to further improve performance on quality benchmark datasets such as LIVE and TID2013. The number of methods is increasing incredibly fast, and not all methods could be included in this study. This also holds true for a further number of more established metrics, such as FSIM \cite{FSIM}, VIF \cite{VIF} and IW-SSIM \cite{IW-SSIM}, that were not included. We especially focused on methods that have been applied to the medical domain more frequently \cite{MacNaughton2023}. The investigation of further methods and their applicability to the medical domain remains further work.
\section{Conclusions}
\label{sec:conclusions}
For the validation of medical image-to-image translation, we gave a broad overview of possible metrics. For 11 reference, 12 non-reference metrics and a segmentation metric, we presented a detailed study of their sensitivity to 11 types of distortions, which are specific for MR images. 
As a conclusion, we give a few recommendations for the selection and application of appropriate validation metrics.
\begin{itemize} 
\item There is no optimal metric. Depending on the application, sensitivity to specific distortions is desired or not. Usually, a combination of metrics should be chosen, considering all expected distortions of the application. As reference metrics, the combination of SSIM, LPIPS, MSE and NMI is able to detect a large set of undesired distortions. MS-SSIM, PCC, DISTS, NMSE, and MAE do not give much additional information. MSE is the most frequently used error metric and therefore best suited for comparison to previous studies. In general, the use of PSNR should not be recommended. CW-SSIM is appropriate in addition to ignore slight misalignments to reference images.
\item MB, BR and NIQE do not yield consistent quality scores for increasing distortion strengths and dramatically deviate for different intensity value ranges. They cannot be recommended for medical images.
\item For detecting blurriness, BE and BEW and CPBD perform very robustly. When images are noisy or show stripe artifacts, BEW is hardly influenced. MB, BR and NIQE do not yield consistent quality scores for increasing distortion strengths and dramatically differ for different normalization methods. They cannot be recommended.
\item Of all tested non-reference metrics, MLC and MSLC are best able to indicate ghosting artifacts.
\item If normalization was used before metric assessment, the method and all relevant parameters must be reported in detail. These parameters must be considered, when comparing scores across studies, because normalization parameters can have a significant effect on absolute metric values.
\item The data range parameter $L$ is commonly used to adapt metrics, that were originally designed for a fixed 8-bit value range, to float valued images with potentially infinite intensity value ranges. However, scaling or binning to the range $(0, 255)$ is often more consistent with the original design due to internal constants  (e.g. SSIM) and thresholds (e.g. JNB). For reference metrics, the data range should be derived at least from both images if not from the whole dataset.
\item For NMI and PCC, a data range parameter $L$ is not needed and normalization can be omitted, if similar intensity values represent similar tissue types. Otherwise, an appropriate normalization method should aim for mapping similar tissue types to similar intensity values. 
\item If source images and target images are not spatially aligned, they must be registered with highest possible precision before evaluating with reference metrics. Rigid translation reduces assessed similarity more substantially than small local elastic deformations. CW-SSIM is robust against small translations. The type of interpolation used for registration may additionally blur the images.
For LPIPS and DISTS, Minmax normalization to [0, 1] and [-1,1] is recommended. For non-quantitative image modalities such as MR, any kind of normalization is always recommended.”
\item A segmentation downstream task is extremely useful for evaluation, because small but relevant structures are assessed. The performance of the segmentation model must be verified before using the segmentations for similarity assessment. 
\end{itemize}
In summary, the metrics for evaluation of image-to-image MR synthesis models must be selected carefully. Frequently used SSIM and PSNR cover a large range of distortions, but have specific weaknesses, that must be covered by other metrics. Specifically PSNR does not seem appropriate for the evaluation of synthetic images.
We suggest to always select metrics, that are able to detect undesired distortions specific and typical for the desired application and that are insensitive towards admitted and expected distortions.
 Which metrics are most appropriate can be directly derived from our experimental relative metric scores.
As metrics are also often used as loss functions for model training \cite{ding2020optim} or validation metrics for model selection, the choice of appropriate metrics can directly improve image synthesis models before clinical validation by human readers and thereby reduce development time and costs. 
\section*{Acknowledments}
The authors like to thank the Bayer team of the AI Innovation Platform for providing compute infrastructure and technical support.
\section*{Author Contributions}
M.D. and M.L. designed the study, M.D. performed the experiments, and analysed the results. All authors reviewed and contributed to the manuscript.
\section*{Additional Information}
\subsection*{Competing interests}
The authors declare no competing interests.
\section*{Data availability}
The BraSyn 2023 dataset is available at \url{www.synapse.org/brats2023}. The code for all metrics and distortions is available at \url{www.github.com/bayer-group/mr-image-metrics}.

\setcitestyle{square}
\bibliography{references}
\newpage

\renewcommand{\thefigure}{S.\arabic{figure}}
\setcounter{figure}{0}
\renewcommand{\thetable}{S.\arabic{table}}
\setcounter{table}{0}

\renewcommand{\thesection}{\Alph{section}}
\setcounter{section}{0}

\section*{
Similarity and Quality Metrics for MR Image-to-Image Translation (Supplement)}
\parbox{\textwidth}{\centering Melanie Dohmen, Mark A. Klemens, Ivo M. Baltruschat, Tuan Truong, Matthias Lenga}
\renewcommand{\thefigure}{S.\arabic{figure}}
\setcounter{figure}{0}
\renewcommand{\thetable}{S.\arabic{table}}
\setcounter{table}{0}
\section{Supplementary Methods}
\subsection{Notation}
\label{sec:notation}
Let $I$ be an image with intensity $I(\mathbf{x})$ at pixel location $\mathbf{x}$. A three-dimensional image $I\in\mathbb{R}^{w\times h \times d}$ of height $h$, width $w$ and depth $d$ consists of $|I| = w \cdot h \cdot d$ pixels, with pixel locations denoted as  $\mathbf{x} = (x_1,x_2,x_3) \in \mathbb{N}^3$ with $0\leq x_1 < h$,  $0 \leq x_2 < w$, $0 \le x_3 < d$. Accordingly, a  two-dimensional image  is defined as $I\in\mathbb{R}^{h\times w}$ with pixel locations $\mathbf{x}= (x_1,x_2)$ and $|I| = h \cdot w$ pixels in total. Let $I_{\textrm{max}}$ denote the maximum intensity, $I_{\textrm{min}}$ the minimum intensity, $\mu_{I}$ the mean intensity, $\sigma_I$ the standard deviation of all image intensities in $I$.
The $k$-th percentile $I_{k\%} \in \mathbb{R}$ of the image $I$ is the smallest intensity value, such that a fraction of $k$\% of all pixels in $I$ have lower or equal intensity value. 
The median of all image intensities in $I$ is $I_{50\%}$. The interquartile range (IQR) is $I_{75\%} -I_{25\%}$.

\subsection{Calculation of Normalization Methods}
\label{sec:normalization_methods_calc}

The calculation for all normalization methods is found here, the notation is detailed in Sec.\,\ref{sec:notation}.

\subsubsection*{Minmax}
Given a (sub-) intensity range $[i_1, i_2]$ the image intensities are shifted and scaled to meet a target intensity range $[j_1, j_2]$. For the default choice of $i_1 = I _{\mathrm{min}}$, $ i_2 = I_{\mathrm{max}}$, $j_1=0$ and $j_2=1$, these parameters are dropped.
\begin{equation}
\begin{aligned}
\mathrm{Minmax}_{[i_1, i_2]\mapsto[j_1, j_2]}(I) = \frac{I - i_1} {i_2 - i_1} \cdot  (j_2 - j_1) + j_1
&\hspace{1cm} &
\mathrm{Minmax}(I) = \frac{I - I_{\textrm{min}} }{I_{\textrm{max}} - I_{\textrm{min}}}\\
\end{aligned}
\end{equation}
In case of constant images with $I_{\textrm{min}} = I_{\textrm{max}}$, the division by zero is omitted, such that the result is a constant image $\mathrm{Minmax}(I) = j_1$.
\subsubsection*{cMinmax}
Clipped Minmax normalization is equal to Minmax normalization with previous clipping at the percentiles $I_{p\%}$ and $I_{q\%}$. If only $p < 50\%$ is given, the default is $q = 100\%-p =95$ and$j_1=0$ and $j_2=1$, which reduces the notation.
\begin{equation}
\begin{aligned}
\mathrm{cMinmax}_
{[I_{p\%}, I_{q\%}]\mapsto[j_1, j_2]}(I) &=\mathrm{Minmax}_{[I_{p\%}, I_{q\%}]\mapsto[j_1, j_2]}((\mathrm{clip}_{I_{p\%},I_{q\%}}(I))  & \hspace{1cm}  &
\mathrm{clip}_{[I_{p\%}, I_{q\%}]}(I) = \begin{cases}
    I_{p\%} &\textrm{, if } I(\mathbf{x}) \le I_{p\%} \\
    I_{q\%} &\textrm{, if } I(\mathbf{x}) \ge I_{q\%} \\
I(\mathbf{x}) &\textrm{, else }
\end{cases} \\
\mathrm{cMinmax}_{p\%}(I) &= \mathrm{Minmax}(\mathrm{clip}_{[I_{p\%},I_{100\%-p\%}]}(I)) \hfill & & \\
\end{aligned}
\end{equation}
In case of (nearly) constant images with $I_{p\%} = I_{q\%}$, the result becomes a constant image with $\mathrm{cMinmax}(I) = j_1$.
\subsubsection*{Zscore} 
\begin{equation}
\mathrm{Zscore}(I)= \frac{I - \mu_{I} }{\sigma_I}
\label{eq:zscore}
\end{equation}
In case of constant images with ${\sigma_I} = 0$, the division by zero is omitted, such that $\mathrm{ZScore}(I) = 0$.
\subsubsection*{Quantile}
\begin{equation}
 \mathrm{Quantile}(I) = 
    \frac{I - I_{50\%}}{I_{75\%}-I_{25\%}}
\end{equation}
In case of a large fraction of equal image values, i.e. $I_{25\%} = I_{75\%}$, the division by zero is omitted, such that $\mathrm{Quantile}(I) = I - I_{50\%}$.
%
%
\subsubsection*{Binning}
The image intensities are mapped to $B$ bin values.
\begin{equation}
 \mathrm{Bin}_B(I) =
 \min(B-1,
\lfloor \mathrm{Minmax}_{[I_{\mathrm{min}}, I_{\mathrm{max}}]\mapsto[0,B-1]} \rfloor ) = \min(B-1,
\lfloor B \cdot(I - I_{\mathrm{min}}) /(I_{\mathrm{max}}-I_{\mathrm{min}}) \rfloor )
\label{eq:binning}
\end{equation}

\newpage

\subsection{Metric Calculation}
\label{sec:metric_calc}
Reference and non-reference metrics are ordered alphabetically. The notation is detailed in Sec.\,\ref{sec:notation}.
\subsubsection*{BE (Blur-Effect)}
For each dimension $d=1,...,n$ of $I$, a blurred version $\tilde{I}$ of image $I$ is created by convolution with a uniform kernel $U_{k,d}$ of size $k$ along $d$. The absolute differences of neighboring pixels along $d$ are in $\tilde{I}$ and $I$ as $\tilde{D}$ and $D$ respectively. $D_d$ and $\tilde{D}_d$ can be seen as the gradients or edge images of the original image $I$ and its blurred version $\tilde{I}$. Then the sum of positive differences $D-\tilde{D}$ is related to the sum of differences $D$ only as a measure of blurriness. BE is implemented in the scikit-image python library \cite{skimage} with a default of $k=11$. In 
detail, we compute for all $d$:
\begin{equation}
\begin{aligned}
& \tilde{I}_d = \mathrm{conv}(U_{k,d}, I) 
& \hspace{1cm}
& ~\\
& D_d = |\nabla_d(I) |
& \hspace{1cm}
& S_d = \sum_{\mathbf{x}} D_d(\mathbf{x})\\  
& \tilde{D}_d  = |\nabla_d(\tilde{I}_d)| 
& \hspace{1cm}
& \tilde{S}_d = \sum_{\mathbf{x}} \max(0,D_d(\mathbf{x}) - \tilde{D}_d(\mathbf{x}))
\end{aligned}
\end{equation}
where $\nabla_d(I) $ denotes the differential of image $I$ in dimension $d$ and $\mathbf{x}$ denotes all pixel locations in $D_d$. 
The final Blur-Effect is defined as
\begin{equation}
\mathrm{BE} 
= \max_{d}\frac{S_d -\tilde{S}_d}{S_d}
\end{equation}
\subsubsection*{BEW (Blurred Edge Widths)}
 First, edge pixels are detected along all dimensions $d$ as $E_d(\mathbf{x}) = 1$ and non-edge pixels as $E_d(\mathbf{x})  = 0$, e.g. with the canny algorithm \cite{canny}. Second, edge pixels are traced along their detected dimension to find the next pixel with a differently signed gradient, which marks the end of the edge. The summed distances of the edge pixel to the ends of the edge determine the edge width $W_d(\mathbf{x})$ for dimension $d$.

Let $d$ be a number of dimensions $\in {0, 1, ..., n}$, then the BEW metric is defined as:
\begin{equation}
\begin{aligned}
&\mathrm{BEW}(\mathbf{x}) = \frac{1}{n} \cdot \sum_d \frac{\sum_{\mathbf{x}} W_d(\mathbf{x})}{\sum_{\mathbf{x}} E_d(\mathbf{x})}
\end{aligned}
\end{equation}
\subsubsection*{BR (Blur Ratio)}
In order to identify edge pixels, the image gradient along dimension $d$ and the mean absolute value $\mu_{D_d}$ are computed:
\begin{equation}
\begin{aligned}
&D_d  = |\nabla_d(I)| 
& \hspace{1cm}
& \mu_{D_d} = \frac{1}{|I|}\sum_{\mathbf{x}} D_d (\mathbf{x})\\
\end{aligned}
\end{equation}
With two further criteria, a binary map of edge pixels $E_d(\mathbf{x}) \in \{0,1\}$ is derived from $D_d$. First, all pixels with gradient values lower or equal to the mean gradient are set to 0 in the edge candidate gradient map $C_d$. Second, the gradient value must exceed the values of its direct neighbors in dimension $d$:
\begin{equation}
\begin{aligned}
&C_d(\mathbf{x}) = \begin{cases}
    D_d(\mathbf{x}) &\textrm{, if } D_d(\mathbf{x}) > \mu_{D_d}\\
    0 &\textrm{, otherwise}
\end{cases}  
&\hspace{1cm}
&E_d(\mathbf{x}) = \begin{cases}
    1 &\textrm{, if } C_d(\mathbf{x}) > C_d(\mathbf{x}_1,...,\mathbf{x}_d+1,..., \mathbf{x}_n)\\
    ~& \textrm{~~and~} C_d(\mathbf{x}) > C_d(\mathbf{x}_1,...,\mathbf{x}_d-1,..., \mathbf{x}_n)\\
    0 &\textrm{, otherwise}
\end{cases} 
\end{aligned}
\end{equation}
In order to determine blurred pixels, the average intensity $A_d(\mathbf{x})$ of the neighbors of $\mathbf{x}$ in dimension $d$ is defined.
\begin{equation}
A_d(\mathbf{x}) =\frac{|I(x_1,..., x_d-1, ..., x_n) - I(x_1, ...,x_d+1, ..., x_n)|}{2}
\end{equation}
And then related to the pixel intensity to define the inverse blurriness $IB_d$ for dimension $d$. A threshold $t_\text{IB}$ identifies blurred pixels, where the inverse blurriness of all dimensions $d$ falls below $t_\text{BR}$.
\begin{equation}
\begin{aligned}
&IB_d(\mathbf{x})\frac{|I(\mathbf{x}) - A_d(\mathbf{x})|}{A_d(\mathbf{x})} 
& \hspace{0.3cm}
&B(\mathbf{x}) = \begin{cases}
    1 &\textrm{, if }\max_{d}(IB_d(\mathbf{x})) < t_\text{IB} \\
    0 &\textrm{, otherwise}
\end{cases}\\
&\mathrm{BR} = \frac{\sum_{\mathbf{x}} B(\mathbf{x})}{\sum_{\mathbf{x}} \max_d(E_d(\mathbf{x}))}&\hspace{1cm}&
&\mathrm{MB} = \frac{\sum_{\mathbf{x}} \max_d IB_d(\mathbf{x})}{\sum_{\mathbf{x}} B(\mathbf{x})}
\end{aligned}
\end{equation}
Based on experiments, the original paper proposes a threshold of $t_\mathrm{IB}=0.1$. The BR metric is defined as the ratio of the total number of blurred pixels and the total number of edge pixels
\subsubsection*{BRISQUE (Blind/Reference-less Image Spatial Quality Evaluator)}
As a first step, the image $I$ is normalized with a modified zscore method\cite{brisque} including a stabilizing constant $C=1$ to $I^{\prime}$, also called mean subtracted contrast normalized (MSCN) coefficients.
\begin{equation}
    I^{\prime} = \frac{I - \hat{\mu}_I}{\hat {\sigma}_I + C}
\end{equation}
In a second step, pairwise products of image intensities in a direct neighborhood are calculated:
\begin{equation}
    \begin{aligned}
    H(I^{\prime}, \mathbf{x}) &= I^{\prime}(\mathbf{x}) \cdot I^{\prime}(x_1+1, x_2) & \text{(horizontal)}\\
    V(I^{\prime}, \mathbf{x}) &= I^{\prime}(\mathbf{x}) \cdot I^{\prime}(x_1, x_2+1) & \text{(vertical)}\\
    D_1(I^{\prime}, \mathbf{x}) &= I^{\prime}(\mathbf{x}) \cdot I^{\prime}(x_1+1, x_2+1) & \text{(first diagonal)}\\
    D_2(I^{\prime}, \mathbf{x}) &= I^{\prime}(\mathbf{x}) \cdot I^{\prime}(x_1-1, x_2+1) & \text{(second diagonal)}\\
    \end{aligned}
\end{equation}
Third, a general Gaussian distribution is fitted to $I^{\prime}$ with parameters shape $s$ and variance $v$ as well as asymmetric Gaussian distributions are fitted to each of $H$, $V$, $D_1$ and $D_2$ with parameters  shape $s$, mean $m$, left variance $lv$, and right variance $rv$, yielding $1 \cdot 2 + 4 \cdot 4 = 18$ parameters.
Finally, the BRISQUE quality score is predicted from these features by a trained support vector regressor.

\subsubsection*{CW-SSIM (Complex-Wavelet Structural Similarity Index Measure}
In the complex wavelet transform domain, suppose
 $\mathbf{c}_{\mathbf{x}, i} = \{c_{\mathbf{x}, i} | i = 1, _N\}$ and $\mathbf{d}_{\mathbf{x}, i} = \{d_{\mathbf{x}, i} | i = 1, _N\}$ are
two sets of coefficients extracted at the same spatial location $\mathbf{x}$ in
the same wavelet subbands of the two images $I$ and $R$ being compared,
respectively. Then, CW-SSIM is defined as
\begin{equation}
    \mathrm{CW-SSIM}(I, R) = \frac{1}{|I|} \sum_{\mathbf{x}\in R}\frac{ 2 \lvert \sum_{i=1}^N c_{\mathbf{x}, i}\cdot d_{\mathbf{x}, i}^{\ast} \rvert + K}{\sum_{i=1}^N |c_{\mathbf{x}, i}|^2 + \sum_{i=1}^N |d_{\mathbf{x}, i}|^2 + K}
\end{equation}
where $d_{\mathbf{x}, i}^{\ast}$ is the complex conjugate of $d_{\mathbf{x}, i}$ and $K$ is a constant to improve robustness, where the local signal-to-noise ratios are low.
For our experiments, we used an implementation\cite{Ding_code} with $K=1\cdot 10^{-12}$ and images were Minmax normalized to range $(0,255)$.

\subsubsection*{CPBD (Cumulative Probability of Blur Detection)}
The probability of detecting blur in a pixel $\mathbf{x}$ of an edge block $\mathit{e}$ can be expressed as an exponential function of $p(\mathbf{x}, \mathit{e})$. Let $\mathcal{X}$ be the set of processed edge pixels, i.e. the set of detected edge pixels located in an edge block $\mathit{e} \in \mathcal{E}$.
Then the CPBD metric corresponds to the cumulative probability of detecting blur in any of the processed edge pixels with a probability $\le 0.63$.
\begin{equation}
\begin{aligned}
p(\mathbf{x}, \mathit{e}) =1 - \exp \left(-  \left| \frac{W_0(\mathbf{x})}{\mathrm{jnb}(\mathit{e}, L)}\right| \right)^{\beta} & \hspace{1cm} &
\mathcal{X} := \left\{\mathbf{x} \mid \max_d E_d(\mathbf{x})=1 \wedge \exists \mathit{e} \in \mathcal{E}: \mathbf{x} \in \mathit{e} \right\}
\end{aligned}
\end{equation}
\begin{equation}
\mathrm{CPBD}(I) = \sum_{\mathbf{x} \in \mathcal{X}}\begin{cases} 1 / |\mathcal{X}| & \text{, if } p(\mathbf{x}, \mathit{e})\le 0.63 \\
0 & \text{, else}
\end{cases}
\end{equation}

We employed canny edge detection with a lower threshold at $0.1\cdot L$ and a higher threshold of $0.2\cdot L$, and scaled $\mathrm{jnb}(\mathit{e})$ with $L$ as defined in Eq.\,(\ref{eq:jnb}).
\subsubsection*{DSC (Dice Similarity Coefficient)} 
 For segmentations $S_I$ and $S_R$ of images $I$ and $R$ respectively, the DSC relates the intersection of both segmentations to the sum of their sizes.
\begin{equation}\label{eq:dice}
        \mathrm{DSC}(S_I, S_R) =  \frac{ 2|S_I \cap R_I| + \epsilon}{|S_I| + |R_I| + \epsilon}
\end{equation}
The small constant $\epsilon > 0$ is typically introduced, to assure that the DSC is not undefined in case both segmentations are empty.
\subsubsection*{DISTS (Deep Image Structure and Texture Similarity)}
Given the feature maps $F_l^i(I)$ and $F_l^i(R)$ from layer $l$ and channel $i$ of a pre-trained network, the structure similarity $S_l^i$ and the texture similarity $T_l^i$ between the feature maps were defined as
\begin{equation}
\begin{aligned}
S_l^i(I, R) = \frac{2 ~ \mu_{F_l^i(I)}\cdot\mu_{F_l^i(R)} + C_1}{\mu_{F_l^i(I)}^2 + \mu_{F_l^i(R)}^2 + C_1} & \hspace{1cm} &
T_l^i(I, R) = \frac{2 ~ \sigma_{F_l^i(I)} \cdot \sigma_{F_l^i(R)} + C_2}{\sigma_{F_l^i(I)}^2 + \sigma_{F_l^i(R)}^2 + C_2}
\end{aligned}
\end{equation}
where $C_1$ and $C_2$ are positive constants.
The overall DISTS metric is obtained by averaging weighted combinations of structure and texture similarities across network layers at different depths levels
\begin{equation}
    \text{DISTS}(I, R) = \sqrt{\sum_{l} \sum_i \left( \alpha_l^i ~ S^i_l(I, R) 
    + \beta_l^i ~ T^i_l(I, R)  \right)^2 }
\end{equation}
where $\alpha_l^i, \beta_l^i$ are optimized weighting factors for each channel $i$ and layer $l$.
\subsubsection*{JNB (Just Noticeable Blur)}
The idea of blurred edge widths (BEW) is extended with a notion of just noticeable blur. The image $I$ is evaluated in smaller blocks $b$ of size $s_b$. Only blocks $\mathit{e} \in \mathcal{E}$ with a sufficient fraction ($T > 0.002$) of edge pixels are considered. The edge widths are additionally weighted with the just noticeable blur width $\mathrm{jnb}(\mathit{e}, L)$ depending on the maximal intensity range per block $e$ and the data range $L$, which is 255 by default for 8-bit images. The exponential parameter $\beta = 3.6$ was also used for adjustments.
\begin{equation}
\begin{aligned}
\mathcal{E}=\left\{b \text{ }\middle| \text{ }\frac{\sum_{\mathbf{x} \in b} \max_d E_d(\mathbf{x})}{s_b} > 0.002\right\} & \hspace{1cm}
& \mathrm{jnb}(\mathit{e}, L)=\begin{cases} 5 & \text{, if }\frac{\mathit{e}_{\mathrm{max}} - \mathit{e}_{\mathrm{min}}}{L} <= \frac{50}{255}\\
3 & \text{, else}
\end{cases}
\end{aligned}
\label{eq:jnb}
\end{equation}
\begin{equation}
\mathrm{JNB}(I) = \frac{1}{|\mathcal{E}|} \cdot \sum_{\mathit{e} \in \mathcal{E}}
    \left(\sum_{\mathbf{x} \in \mathit{e}} \left( \frac{W_0(\mathbf{x})}{\mathrm{jnb}(\mathit{e}, L)}\right)^{\beta} \right)^{1 / \beta} 
\end{equation}
\subsubsection*{LPIPS (Learned Perceptual Image Patch Similarity)} 
The LPIPS metric extracts feature maps $F_l^i(I)$ and $F_l^i(R)$ from layer $l$ and channel $i$ of the pre-trained network. The feature maps are unit-scaled to $\hat{F}_l^i(I), \hat{F}_l^i(R)$, weighted channel-wise with vectors $w_l$ and subtracted. Then  the corresponding $L_2$ norms are averaged spatially and summed to obtain the LPIPS distance $\mathrm{LPIPS}(I,R)$:
\begin{equation}
\mathrm{LPIPS}(I, R) 
=  \sum_l \frac{1}{|F_l|}\sum_{\mathbf{x}_l}\left| w_l \odot (\hat{F}_l(I) -  \hat{F}_l(R))(\mathbf{x}_l) \right|_2^2 
\end{equation}
where $\mathbf{x}_l$ denotes a pixel location in feature map $F_l$, $\odot$ denotes the channel-wise multiplication and $|\cdot|_2$ denotes the euclidean norm.
The linear weights $w_l^i$ were optimized with the Berkeley-Adobe Perceptual Patch Similarity
(BAPPS) Dataset \cite{lpips}. 
\subsubsection*{MAE (Mean Absolute Error)} 
\begin{equation}
\textrm{MAE}(I, R) = \frac{1}{|I|}\sum_{\mathbf{x}} |R(\mathbf{x}) - I(\mathbf{x})| 
\end{equation}
\subsubsection*{MB (Mean Blur)}
See the BR metric for details on how to compute the blurred edges $B$ and the inverse blurriness $IB$.
The MB metric is defined as the ratio of the summed inverse blurriness to the number of blurred pixels:
\begin{equation}
\mathrm{MB} = \frac{\sum_{\mathbf{x}} \max_d IB_d(\mathbf{x})}{\sum_{\mathbf{x}} B(\mathbf{x})}
\end{equation}
\subsubsection*{MLC (Mean Line Correlation)}
The mean line correlation (MLC) metric is defined as the mean correlation between directly neighbored image lines:
\begin{equation}
\mathrm{MLC} = \frac{1}{w} \sum_{x=1}^{w} \mathrm{PCC}\left(I_c(x), I_c(x+1)\right) + \frac{1}{h} \sum_{y=1}^{h} \mathrm{PCC}\left(I_r(y), I_r(y+1)\right) 
\end{equation}
\subsubsection*{MSE (Mean Squared Error)} 
\begin{equation}
\textrm{MSE}(I, R) = \frac{1}{|I|}\sum_{\mathbf{x}} (R(\mathbf{x}) - I(\mathbf{x}))^2 
\end{equation}
\subsubsection*{MSLC (Mean Shifted Line Correlation)}
The mean shifted line correlation (MSLC) metric is defined as the mean correlation between image lines, that are separated by $\lfloor w/2\rfloor$ or $\lfloor h/2\rfloor$ respectively.
\begin{equation}
\mathrm{MLC} = \frac{1}{\lfloor w/2\rfloor} \sum_{x=1}^{\lfloor w/2\rfloor} \mathrm{PCC}\left(I_c(x), I_c(x+1)\right) + \frac{1}{\lfloor h/2\rfloor} \sum_{y=1}^{\lfloor h/2\rfloor} \mathrm{PCC}\left(I_r(y), I_r(y+1)\right) 
\end{equation}
\subsubsection*{MS-SSIM (Multi-Scale Structural Similarity Index Measure)}
As for SSIM, luminance, contrast and structure are defined locally. For MS-SSIM they are calculated on different scales. 
Let $I_M$ be the M-times low-pass filtered and downscaled version of image $I$. Then MS-SSIM is calculated locally as
\begin{equation}
\mathrm{local-MS-SSIM}(I, R, \mathbf{x}) = l(I_M, R_M, \mathbf{x})^{\alpha_M} \cdot \prod_{j=1}^M c(I_j, R_j, \mathbf{x})^{\beta_j} \cdot s(I_j, R_j, \mathbf{x})^{\gamma_j}  
\end{equation}
where the weighting factors $\alpha_j = \beta_j = \gamma_j$ were determined experimentally for $j=1\dots 5$ as $[0.0448, 0.2856, 0.3001, 0.2363, 0.1333]$.
Then the overall MS-SSIM value is averaged over all pixel locations:
\begin{equation}
\mathrm{MS-SSIM}(I, R) = \frac{1}{|I|} \sum_{\mathbf{x} \in R} l(I_M, R_M, \mathbf{x})^{\alpha_M} \cdot \prod_{j=1}^M c(I_j, R_j, \mathbf{x})^{\beta_j} \cdot s(I_j, R_j, \mathbf{x})^{\gamma_j}  
\end{equation}
\subsubsection*{MTV (Mean Total Variation)}
\begin{equation}
    \mathrm{MTV}(I) = \frac{1}{|I|} \sum_{\mathbf{x}}\sqrt{\sum_d \left( I(\mathbf{x}) - I(x_1, ..., x_d +1, .., x_n) \right)^2}
\end{equation}
\subsubsection*{NIQE (Natural Image Quality Evaluator)}
The same 18 features are extracted as in the calculation of the BRISQUE metric. Additionally, The features are collected again on the image downscaled by factor two to obtain 36 features in total. This is done for all patches of size $96 \times 96$ in the image to be tested, and it was previously done for patches of reference training set consisting of undistorted natural images. The distribution of these features $f_1, \dots, f_{36}$  can be fitted to a multivariate Gaussian model with mean $\nu$ and covariance matrix $\Sigma$. From patches of the reference dataset, $\nu_R$ and $\Sigma_R$ were obtained. For any test image $I$, the fitted model parameters are denoted $\nu_I$ and $\Sigma_I$. The NIQE score is then defined as the difference between the fitted model parameters:
\begin{equation}
    \mathrm{NIQE}(I) = \sqrt{\left( (\nu_R - \nu)^{-T} \cdot (\frac{\Sigma_R - \Sigma_I}{2})^{-1} \cdot (\nu_R - \nu_I) \right)}
\end{equation}

\subsubsection*{NMI (Normalized Mutual Information)}
The joint probability distribution 
$p(I=i,R=r)$ represents the likelihood of pairs of intensity values $i = I(\mathbf{x})$ and $r = R(\mathbf{x})$ occurring at any pixel location $\mathbf{x}$ in $I$ and $R$. The joint probability distribution is computed by counting how many pixel locations $x$ in $I$ and $R$ and dividing this number by the number of pixels $|I|$.
Given an intensity $i\in [0,b-1]$, the probability $p(I=i)$  is computed as the number of pixel locations $\mathbf{x}$ with $I(\mathbf{x}) = i$ divided by the total number of pixels:
\begin{equation}
    \begin{aligned}
        p(I=i) 
        = \frac{ |\{\mathbf{x} ~|~ I(\mathbf{x}) = i\}| }{|I|}  
        & \hspace{5cm} 
        & p(R=r) = \frac{|\{\mathbf{x} ~|~ R(\mathbf{x}) = r\}|}{|R|}
    \end{aligned}
\end{equation}
The Mutual information MI is then defined as
\begin{equation}
\mathrm{MI}(I, R) = \sum_{i, r} p(I=i, R=r) \log \left( \frac{p(I=i, R=r)}{p(I=i) p(R=r)} \right)
\end{equation}
where the sum is taken over all possible intensity values $i,r\in [0,b-1]$ in the images $I$ and $R$. MI can be expressed in terms of the entropy $\mathrm{H}(I)$, $\mathrm{H}(R)$ and $\mathrm{H}(I,R)$ related the distributions $p(I)$, $p(R)$ and $p(I,R)$, respectively. We have:
\begin{equation}
\begin{aligned}
& \mathrm{H}(I) = -\sum_{i} p(I=i)\log p(I=i)
& \hspace{2cm} 
&\mathrm{H}(R) = -\sum_{r} p(R=r)\log p(R=r)\\
&\mathrm{H}(I,R) = -\sum_{i,r} p(I=i, R=r)\log p(I=i, R=r)
& \hspace{2cm} 
& \mathrm{MI}(I, R) = \mathrm{H}(I) + \mathrm{H}(R) - \mathrm{H}(I,R)
\end{aligned}
\end{equation}
The normalized mutual information (NMI) is defined as follows:
\begin{equation}
\begin{aligned}
    \mathrm{NMI}(I, R) &= \frac{\mathrm{MI}(I,R)}{\mathrm{H}(I,R)} + 1 = \frac{\mathrm{H}(I) + \mathrm{H}(R)}{\mathrm{H}(I,R)}
\end{aligned}
    \end{equation}
\subsubsection*{NMSE (Normalized Mean Squared Error)} 
For images $I$ and $R$, the normalized mean squared error is defined as
\begin{equation}
\textrm{NMSE}(I, R) = \frac{1}{|I| \cdot \sigma_R} \cdot \sum_{\mathbf{x}} (R(\mathbf{x}) - I(\mathbf{x}))^2
\end{equation}
where $\sigma_R$ denotes the corrected sample standard deviation of the intensity value distribution of image $R$.
\subsubsection*{PCC (Pearson Correlation Coefficient)}
\begin{equation}
\mathrm{PCC}(I,R) = \frac{\sum _{\mathbf{x}}(I(\mathbf{x}) - \mu_I)(R(\mathbf{x}) - \mu_R)}{\sqrt{\sum _{\mathbf{x}}(I(\mathbf{x}) - \mu_I)^2} \sqrt{\sum _{\mathbf{x}}(R(\mathbf{x}) - \mu_R)^2}}
\end{equation}
where the summations are taken over all pixel locations $\mathbf{x}$ and $R(\mathbf{x})$, $I(\mathbf{x})$ denote the respective intensity values at that location. 
\subsubsection*{PSNR (Peak Signal-to-Noise Ratio)}
Please see MSE for the mean squared error.
\begin{equation}
\mathrm{PSNR}(I, R) 
= 10 \cdot \log_{10}\left(\frac{L^2}{\mathrm {MSE}(I,R)}\right)
= 20 \cdot \log_{10} ( L ) - \log_{10}(\mathrm{MSE}(I,R) )
\label{eq:PSNR}
\end{equation}
\subsubsection*{RMSE (Root Mean Squared Error)}
\begin{equation}
\textrm{RMSE}(I, R) = \sqrt{\frac{1}{|I|}\sum_{\mathbf{x}} (R(\mathbf{x}) - I(\mathbf{x}))^2}
\end{equation}
\subsubsection*{SSIM (Structural Similarity Index Measure)}
The structural similarity index measure (SSIM) combines local image luminance, contrast, and structure. Mean, standard deviation and covariance are calculated locally for each pixel location $\mathbf{x}$ within a $d$-dimensional Gaussian kernel of size $11$ and $\sigma = 1.5$ and are denoted by $\mu_{I}(\mathbf{x})$, $\sigma_I(\mathbf{x})$, $\mu_{R}(\mathbf{x})$, $\sigma_R(\mathbf{x})$ and $\sigma_{I,R}(\mathbf{x})$ respectively. For each pixel location $\mathbf{x}$ and its local neighborhood in images $I$ and $R$, the luminance $l$, contrast $c$ and structure $s$ are defined as:
\begin{equation}
\begin{aligned}
\mathrm{l}(I, R, \mathbf{x})& = \left(\frac{2\sigma_{I}(\mathbf{x}) \sigma_{R}(\mathbf{x})+C_{2}}{\sigma_{I}(\mathbf{x})^{2}+\sigma_{R}(\mathbf{x})^{2}+C_{2}}\right)^\alpha\\
\mathrm{c}(I, R, \mathbf{x}) &=\left(
\frac{2\mu_{I}(\mathbf{x})\mu_{R}(\mathbf{x})+C_{1}}{ \mu_{I}(\mathbf{x})^{2}+\mu_{R}(\mathbf{x})^{2}+C_{1}}\right)^\beta \\
 \mathrm{s}(I, R, \mathbf{x}) &= \left(\frac{\sigma_{I, R}(\mathbf{x})+C_{3}}{ \sigma_{I}(\mathbf{x})\sigma_{R}(\mathbf{x})+C_{3}}\right)^\gamma \\
\end{aligned}
\label{eq:ssim}
\end{equation}   
where $C_1$, $C_2$, $C_3$ are constants to avoid division by arbitrarily small numbers.
The local luminance, contrast and structure are then multiplied and averaged for all pixel locations $\mathbf{x}$. Commonly, structure, luminance and contrast terms are weighted equally with $\alpha=\beta=\gamma = 1$. Choosing in addition $C_3=C_2/2$ yields a simplified formula:
\begin{equation}
\textrm{SSIM}(I, R) = \frac{1}{|I|}\sum_{\mathbf{x} \in R}
\frac{(2 \mu_I(\mathbf{x}) \mu_R(\mathbf{x}) + C_1) 
\cdot(2\sigma_{I,R}(\mathbf{x}) + C_2)}{(\mu_I(\mathbf{x})^2+ \mu_R(\mathbf{x})^2 + C_1) + 
(\sigma_I(\mathbf{x})^2 + \sigma_R(\mathbf{x})^2 + C_2)}  
\end{equation}
The other constants are commonly selected as $C_1=(0.01\cdot L)^2$ and $C_2=(0.03 \cdot L)^2$ with $L = 255$ defined as the data range of the intensity values, in the common case of 8-bit unsigned integers,  $L=255$\cite{ssim}, otherwise $L =\textrm{max}(I_{\mathrm{max}}, R_{\mathrm{max}}) - \textrm{min}(I_{\mathrm{min}}, R_{\mathrm{min}})$.
See also CW-SSIM and MS-SSIM.
\subsubsection*{VL (Variance of Laplacian)}
The Laplacian of an image $I$ can be calculated by convolution ($\ast$) with the Laplacian filter:
\begin{equation}
\mathrm{Laplace}(I) = I \ast \left(
\begin{array}{ccc}
    0 & 1 & 0\\
    1 & -4 & 1\\
    0 & 1 &  0\\
\end{array} \right)
\end{equation}
Then, the variance is calculated:
\begin{equation}
\mathrm{VL}(I) = \frac{1}{|I|}\sum_{\mathbf{x} \in I} \left(\mathrm{Laplace}(I)(\mathbf{x}) - \mu_{\mathrm{Laplace}(I)}\right)^2
\end{equation}
%
%
\subsection{Calculation of Distortions}
\label{sec:distortion_calc}
For each type of distortion, the range of parameters is given for strengths $s=1$ to $s=5$. All other strengths are interpolated linearly between the given values. In addition to the notation in Sec.\,\ref{sec:notation}., the distorted image is denoted as $\tilde{I}$. Let $\mathbf{s}=(w, h)$ be the image size vector of $I$. The minimum and maximum intensity of $I$ are denoted by $I_{\mathrm{min}}$, $I_{\mathrm{max}}$. Inverted transformations, that can restore a transformed image, are denoted by $^{-1}$, e.g. $\mathrm{Minmax}^{-1}$, $\mathrm{FFT}^{-1}$ or $\mathrm{shift}^{-1}$. Distortions are ordered alphabetically.

\subsubsection*{Bias Field,  $c \in [0.5, 10]$}

\begin{minipage}{0.45\textwidth}
An artificial bias field is created from the polynomial function $P_3$ and applied to the image:
\begin{eqnarray}
P_3(x_1, x_2) &=& 10x_1^2(x_1-1) \cdot (x_2-0.5)x_2(x_2-1) \hspace{1cm}\\
\tilde{I} &=& I \cdot e^{c \cdot P_3}
\end{eqnarray}
\subsubsection*{Elastic Deform, $n \in[18, 11], d\in[0.03, 0.1]$}
A grid with 
$n^2$ points is created, and points are displaced by a displacement vector sampled from $\mathcal{N}(\mu=\mathbf{0}; \sigma= d \cdot \frac{\mathbf{s}}{n})$. The image is interpolated along the grid.
\end{minipage}
\begin{minipage}{0.45\textwidth}
\centering
\includegraphics[width=4cm, clip, trim=1cm 1.2cm 1cm 1.2cm ]{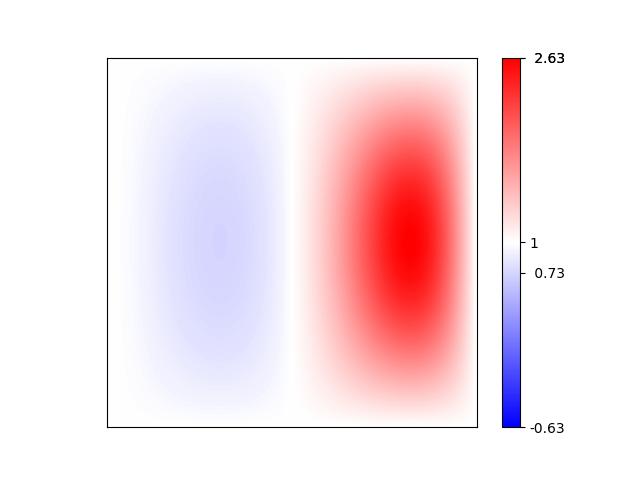}\\
$P_3(x_1, x_2)$
\end{minipage}
\subsubsection*{Gamma Low / Gamma High, $\log(\gamma) \in[-0.01, -0.916] $ / $\log(\gamma) \in[0.095, 0.916]$ }
\begin{equation}
\tilde{I}= \mathrm{Minmax}^{-1}(\mathrm{Minmax}(I)^{\gamma})
\end{equation}
\subsubsection*{Gaussian Blur, $\sigma \in [0.2, 1.3]$}
The image is convoluted with a Gaussian kernel $G_{\sigma}$:
\begin{equation}
\tilde{I}= \mathrm{Conv}(G_{\sigma}, I)
\end{equation}
\subsubsection*{Gaussian Noise, $\sigma \in [0.005, 0.05]$}
\begin{equation}
\tilde{I}= I + \mathcal{N}(\mu=0;\sigma)
\end{equation}
\subsubsection*{Ghosting,  $i \in [0.05, 0.4]$}
The fast Fourier transform (FFT) transforms the image to the frequency domain. The shift operator then moves the center of the transformed image to $0$. The spectrum is scaled (distorted) at every second pixel (to generate two ghosts) along the first axis, then the image is restored for all pixels with $x_1 = w/2$.

\begin{minipage}{0.45\textwidth}
\begin{eqnarray}
I^\prime &=& \mathrm{shift}(\mathrm{FFT}(I))\\
I^{\prime\prime}(x_1, x_2)& =& I^{\prime} \cdot i \textrm{, for } x_1 \mathbin{\%} 2 = 0
\end{eqnarray}
 \end{minipage}
\begin{minipage}{0.45\textwidth}
\begin{eqnarray}
I^{\prime\prime}(x_1, x_2)& = &I^{\prime} \textrm{, for } x_1 = w/2 \\
\tilde{I} &=& \mathrm{shift}^{-1}(\mathrm{FFT}^{-1}(I^{\prime\prime}))
\end{eqnarray}
 \end{minipage}
\subsubsection*{Replace Artifact, $f \in[0.1, 1.0]$}
Copy and mirror a fraction of the upper half of the image to the lower half:
\begin{equation}
\tilde{I}(x_1,x_2)= \begin{cases} I(x_1,x_2) & \textrm{, if }x_2 \le \frac{h}{2} \\ I(x_1, h-x_2)& \textrm{, if } x_2 > \frac{h \cdot (1 + f)}{2}  \end{cases}
\end{equation}
\subsubsection*{Shift Intensity, $f \in[0.05, 0.25]$}
\begin{equation}
\tilde{I}= I + f \cdot (I_{\mathrm{max}}-I_{\mathrm{min}})
\end{equation}
\subsubsection*{Stripe Artifact, $i \in[0.05, 0.5]$}
The fast Fourier transform (FFT) transforms the image to the frequency domain. The shift operator then moves the center of the transformed image to $0$. The spectrum is scaled (distorted) at a single pixel $(0.3\cdot \cos(0), 0.3 \cdot \sin(0)$. Then the image is transformed back to the spatial domain again.
\begin{equation}
\begin{aligned}
    &I^\prime =  \mathrm{FFT}(\mathrm{shift}(I)) \\
    &I^{\prime\prime}(x_1, x_2) = i \cdot I_{\mathrm{max}}\textrm{, for }x_1 = 0.3 \cdot \cos{(0)} \textrm{, } x_2 = 0.3 \cdot \sin{(0)} \\ 
    &\tilde{I} =\mathrm{FFT}^{-1}(\mathrm{shift}^{-1}(I^{\prime\prime})) 
\end{aligned}
\end{equation}
\subsubsection*{Translation, $f \in [0.01, 0.2]$}
\begin{equation}
    \tilde{I}(x_1, x_2)= I (f \cdot w + x_1, f \cdot h + x_2)
\end{equation}
\subsection{Reader study for distortion parameterization}
For finding reasonable and comparable distortion parameter ranges, a reader study was conducted. Six researchers with at least three years experience in the field of medical image anaylsis were asked to partition a sequence of images with increasing distortion parameter settings into four categories:
\begin{enumerate}
\item distortions not visible compared to the reference image, perfect quality
\item visible distortions, but irrelevant difference to the reference, very good quality
\item barely acceptable distortions, could be useful for certain diagnostic questions, poor quality
\item excessive distortions, images cannot be used for any diagnostic question
\end{enumerate}
All distorted image versions belonging to one reference image were shown at a time, in a scrollable table of nine distortions and 20 distortion parameter settings.
For 20 reference images (20 T1-weighted images of the BraTS dataset, 10 of them contrast enhanced), the proposed categorization was evaluated. Additionally, comments of the readers were possible, in case all presented images were perceived outside of category 1 or 4.

In the reader study, all images were normalized to an 8-bit range by Minmax normalization. The distortions were applied to this range and then clipped to the (0, 255) range. This was done in order to visualize and overview the entire intensity value range of the reference and distorted images. The median of the strongest distortion parameter setting, that was assigned to category 1,  $s_{\mathrm{min}}$, was later used to find a parameter setting for strength = 1. The median of the weakest distortion parameter setting, that was assigned to category 4, $s_{\mathrm{max}}$, was later used to find a parameter setting for strength 5.

\begin{figure}
\centering
\includegraphics[width=0.75\textwidth]{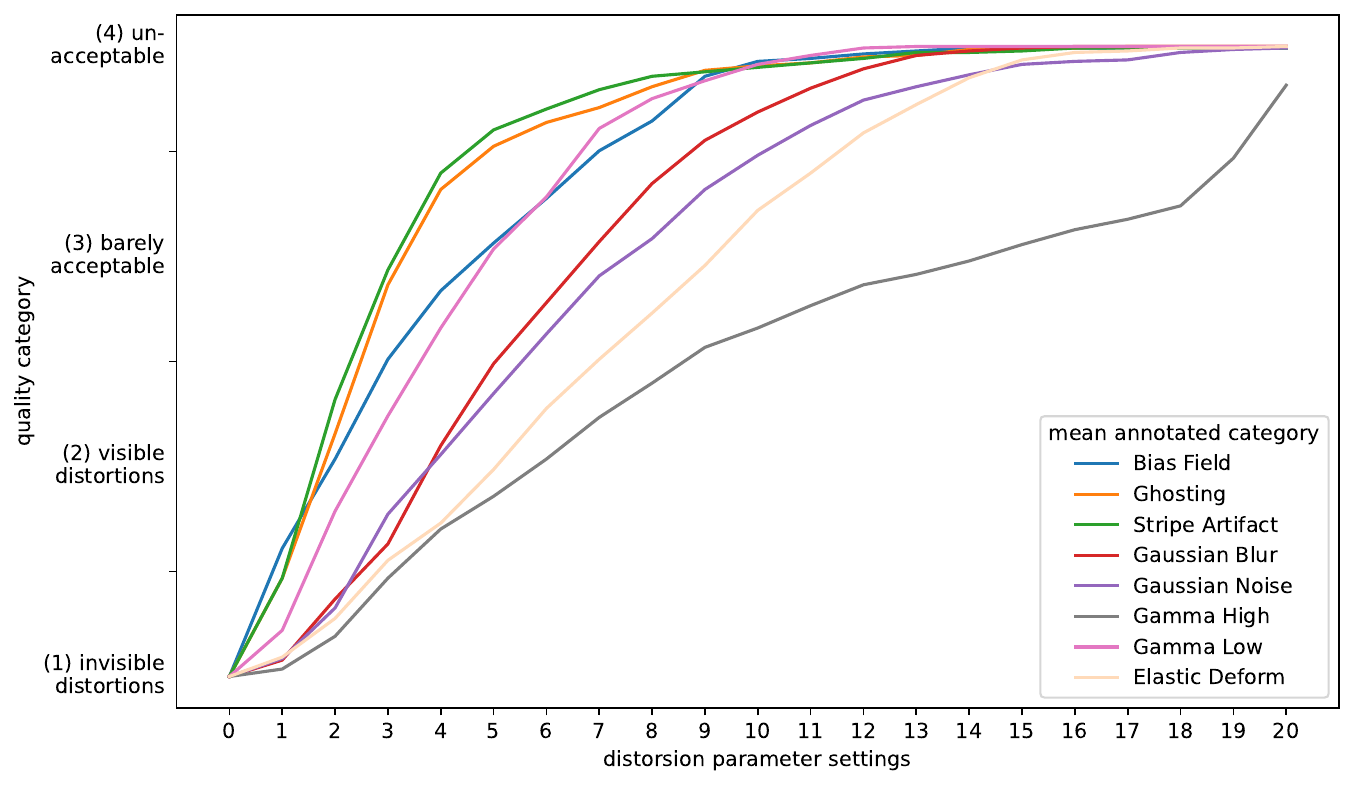}
\caption{Mean category assigned by the six readers for 20 initial parameter settings for each distortion.}
\label{fig:readerstudy}
\end{figure}

As a result from the reader study (see Fig.~\ref{fig:readerstudy}), including collected comments, the initial parameter ranges for Gamma High and Ghosting were extended, in order to better cover the desired quality categories.
For the final experiments, further minor adaptions were made to the distortions to guarantee non-varying stripes, and an equally darkening and brightening bias field. In most cases, distortion intensity parameters were additionally multiplied with the image intensity value range, allowing to apply distortions to the original full MR image intensity range. This allows our analysis to evaluate reference and distorted images with their full intensity range without information loss. Furthermore, the effect of different normalization methods can be investigated. 

The parameterization of the investigated distortions was initialized from results of the reader study on images reduced to 8-bit. The transfer of these parameters to images with the original MR intensity range was estimated by the authors. Directly deriving the parameters from the reader study would impose reducing the intensity value ranges of the images being inspected to a visible range. How this can be solved is ongoing research and has been questioned, also in \cite{kastryulin}.
%
\section{Supplementary Figures}
\subsection{Sensitivity to Strength of Spatial distortions}
\label{sec:metrics_strengths_results}
 Metric scores for different distortion strengths of Translation and Elastic Deform were compared. All metrics scores shown here were assessed on images without normalization, except LPIPS, DISTS and DSC. $^*$: For LPIPS Minmax normalization to [-1, 1] was used, for DISTS Minmax  normalization to [0, 1] was applied and DSC was assessed after Zscore normalization and segmentation.

\begin{tabular}{p{2.2cm}crrrrrrrrrrrr}
\hspace{0.5cm} metrics \rotatebox[origin=c]{90}{\parbox{1.2cm}{\hspace{1.2cm}}} & &\multirow[c]{2}{*}{\metricboxup{SSIM}}&\multirow[c]{2}{*}{\metricboxup{MS-SSIM}}&\multirow[c]{2}{*}{\metricboxup{CW-SSIM}}&\multirow[c]{2}{*}{\metricboxup{PSNR}}&\multirow[c]{2}{*}{\metricboxdown{MSE}}&\multirow[c]{2}{*}{\metricboxdown{NMSE}}&\multirow[c]{2}{*}{\metricboxdown{MAE}}&\multirow[c]{2}{*}{\metricboxdown{LPIPS$^*$}}&\multirow[c]{2}{*}{\metricboxdown{DISTS$^*$}}&\multirow[c]{2}{*}{\metricboxup{NMI}}&\multirow[c]{2}{*}{\metricboxup{PCC}}&\multirow[c]{2}{*}{\metricboxup{DSC$^*$}}\\
Distortions &$s$&&&&&&&&&&&\\
\hline
Reference  & 0&1.00&1.00&1.00&$\infty$&0.00&0.00&0.00&0.00&0.00&2.00&1.00&1.00\\
Translation& 1&0.83&0.87&0.98&26.39&143.10& 2.13$\cdot 10^{5}$&186.02&0.07&0.08&1.27&0.92&0.83\\
Elastic Deform & 5&0.88&0.92&0.91&28.24&105.88& 1.30$\cdot 10^{5}$&116.73&0.06&0.11&1.32&0.95&0.86\\
\end{tabular}
\subsection{Examples of Distorted Images}
In this section we provide more examples of distorted images for all distortions and all strengths for the reader to get an impression about the distortion strengths on different images. The images are visualized with differently intensity ranges, specified in the captions.

\begin{figure}[tbh]
\centering
\includegraphics[width=0.85\textwidth]{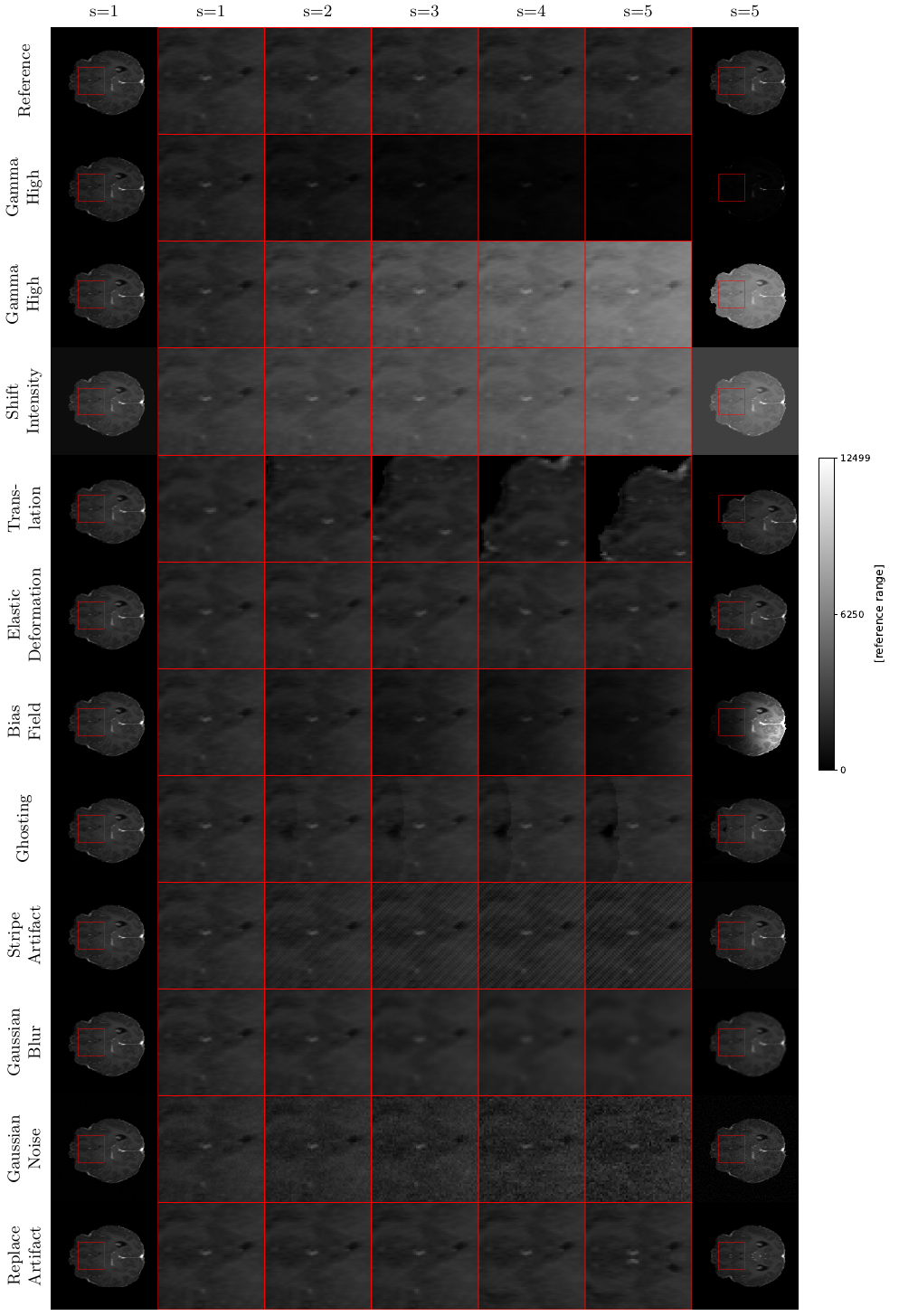}
\caption{Distorted versions of BraTS-GLI-00005-000-t1c, visualized with the intensity range of the reference image.}
\label{fig:distorted_examples_case00005-000_reference}
\end{figure}

\begin{figure}[tbh]
\centering
\includegraphics[width=0.85\textwidth]{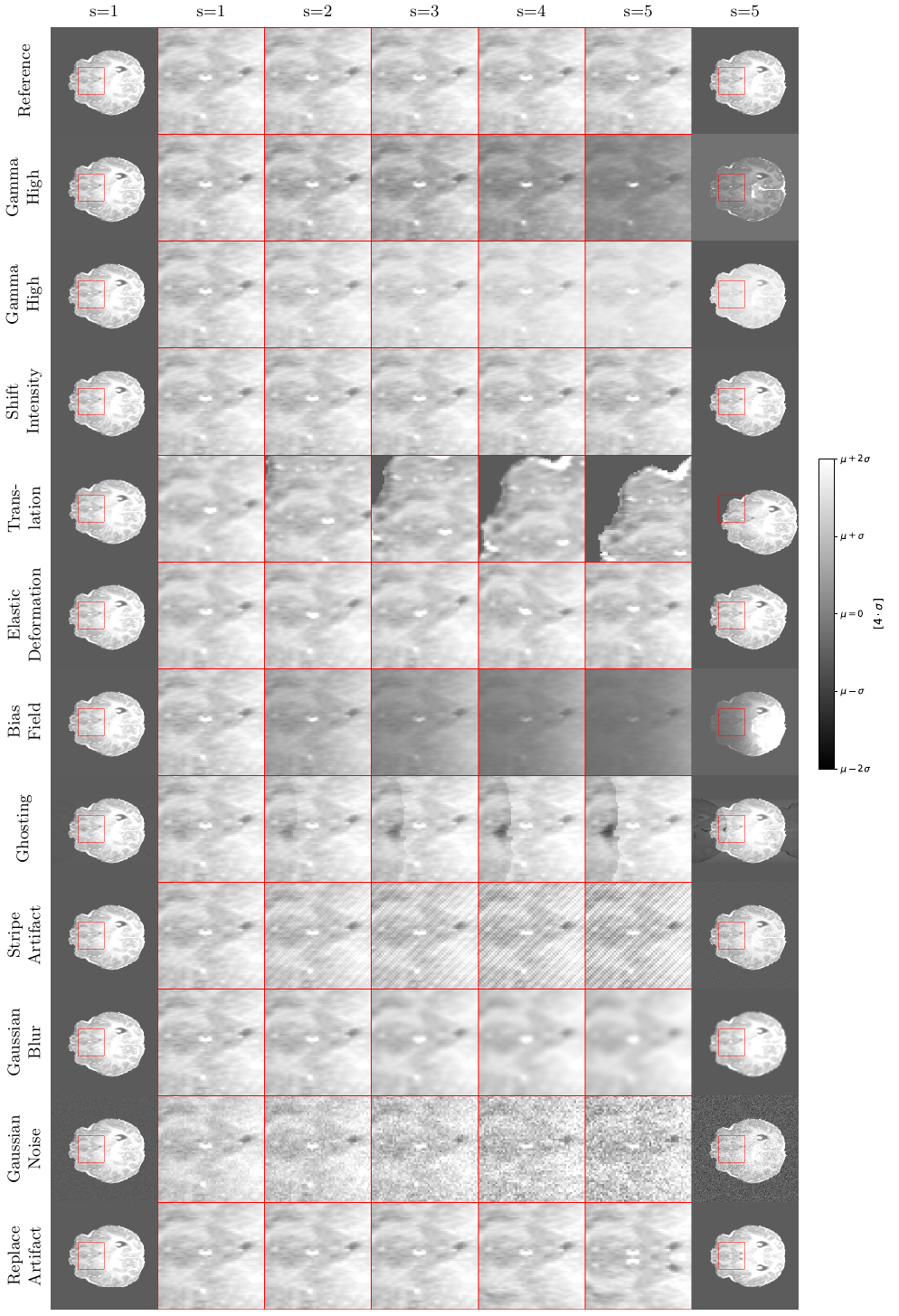}
\caption{Distorted versions of BraTS-GLI-00000-005-t1c, visualized with an intensity range of four standard deviations around the mean of each image individually.}
\label{fig:distorted_examples_case00005-000_zscore}
\end{figure}

\begin{figure}[tbh]
\centering
\includegraphics[width=0.85\textwidth]{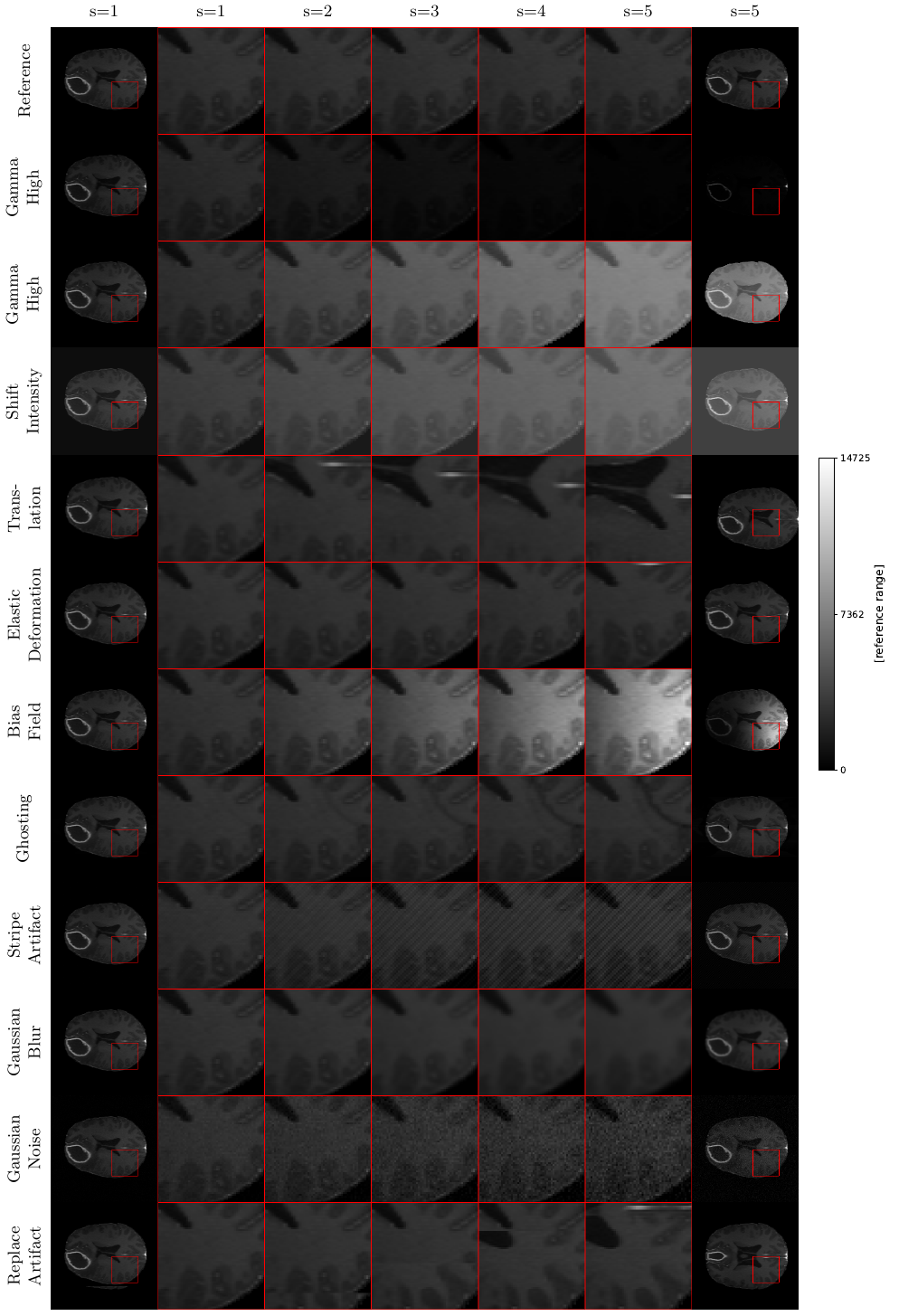}
\caption{Distorted versions of BraTS-GLI-00006-000-t1c, visualized with the intensity range of the reference image.}
\label{fig:distorted_examples_case00006-000_ref}
\end{figure}

\begin{figure}[tbh]
\centering
\includegraphics[width=0.85\textwidth]{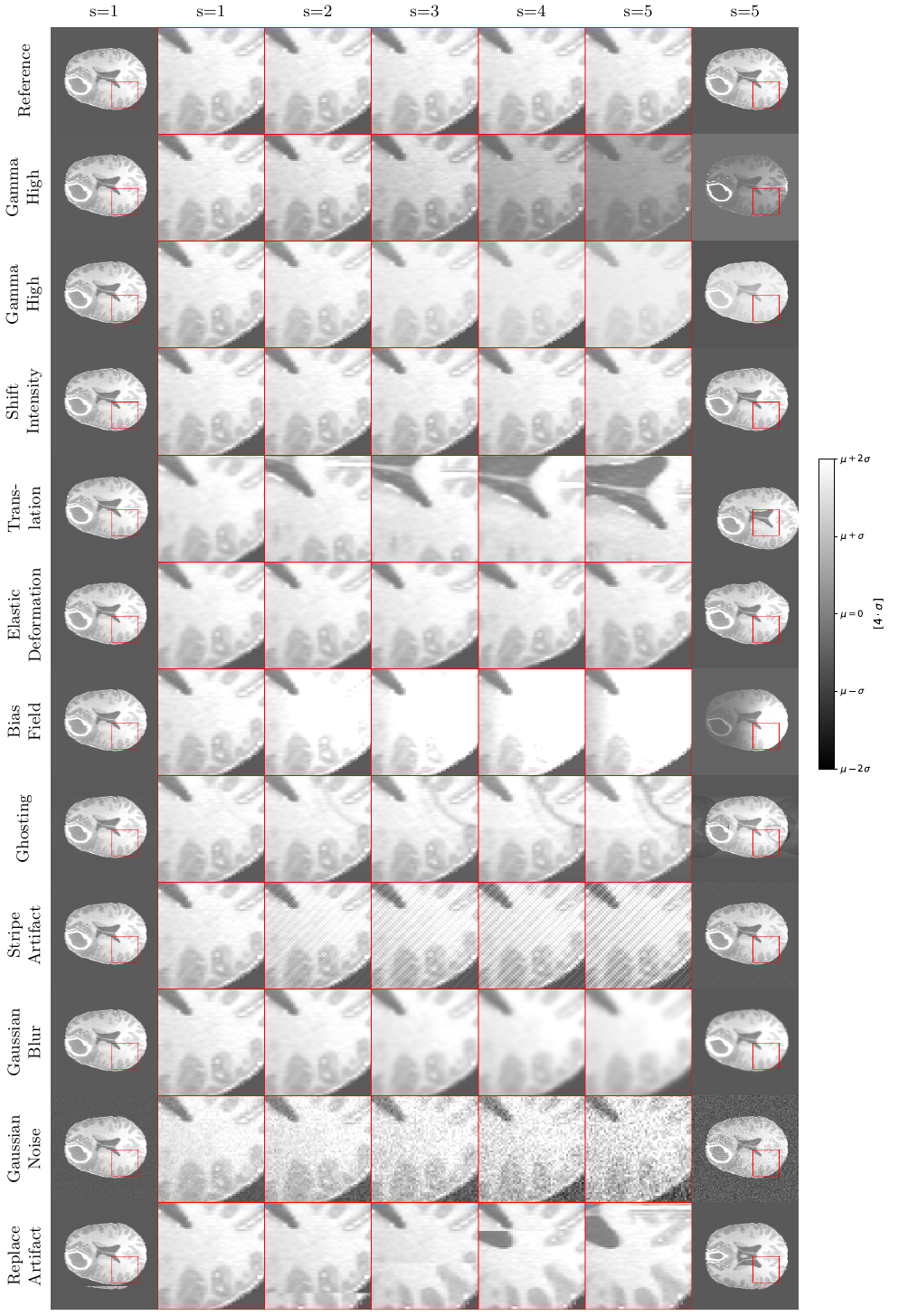}
\caption{Distorted versions of BraTS-GLI-00006-000-t1c, visualized with an intensity range of four standard deviations around the mean of each image individually.}
\label{fig:distorted_examples_case00006-000_zscore}
\end{figure}

\begin{figure}[tbh]
\centering
\includegraphics[width=0.85\textwidth]{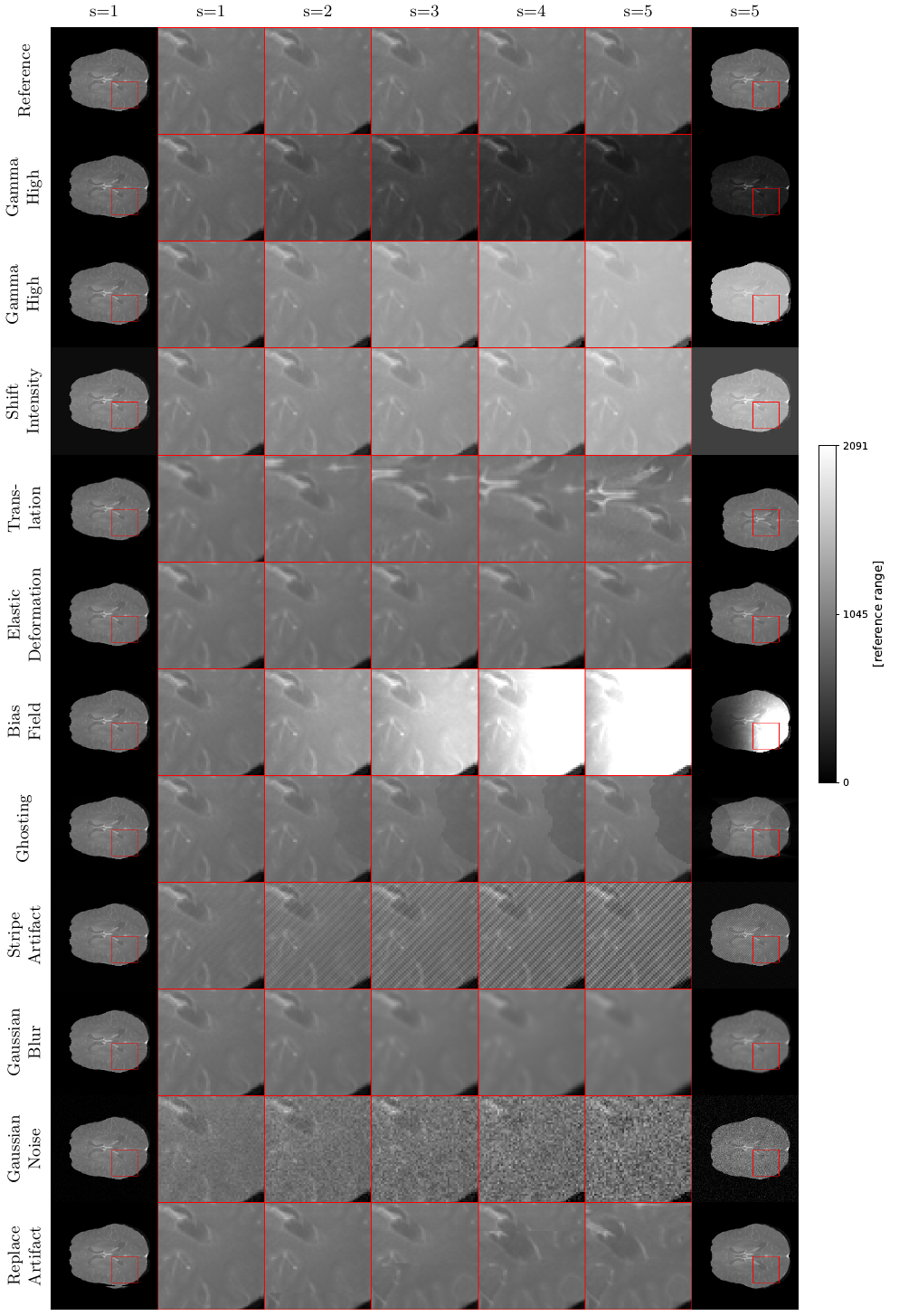}
\caption{Distorted versions of BraTS-GLI-00014-001-t1c, visualized with the intensity range of the reference image.}
\label{fig:distorted_examples_case00014-001_ref}
\end{figure}

\begin{figure}[tbh]
\centering
\includegraphics[width=0.85\textwidth]{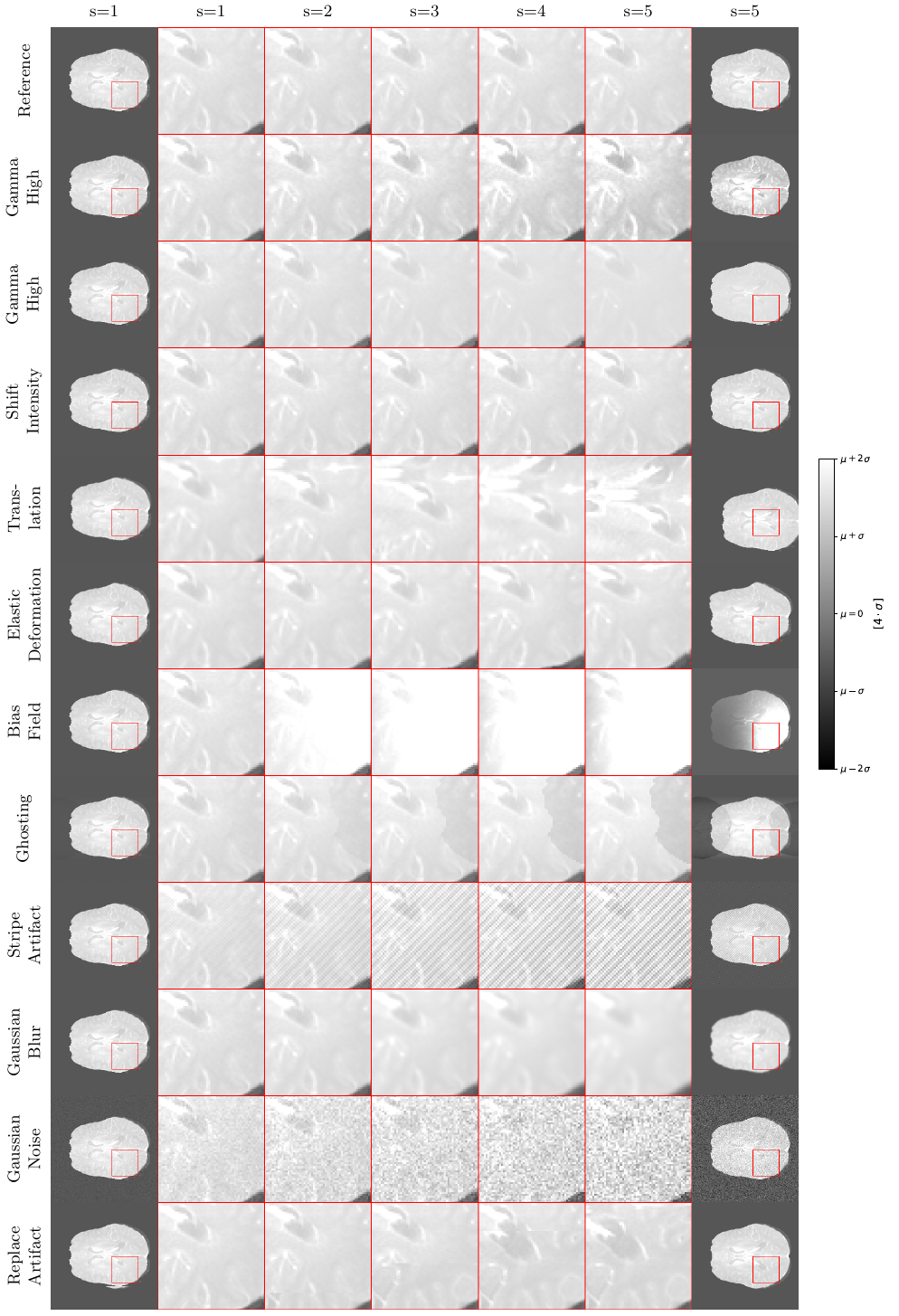}
\caption{Distorted versions of BraTS-GLI-00014-001-t1c, visualized with an intensity range of four standard deviations around the mean of each image individually.}
\label{fig:distorted_examples_case00014-001_zscore}
\end{figure}

\clearpage
\subsection{Evaluation Plots for All Reference Metrics and Normalization Methods}
\label{sec:metrics_evaluation_plots}
In the following figures we present six  plots for each combination of reference metric, one for each normalization method, including w/o normalization. For each distortion and strength the median of all metric scores across all 100 cases is plotted, such that an increasing or decreasing trend is usually observed along the distortion strengths.

\begin{figure}[htb]%
\centering
\includegraphics[width=\textwidth]{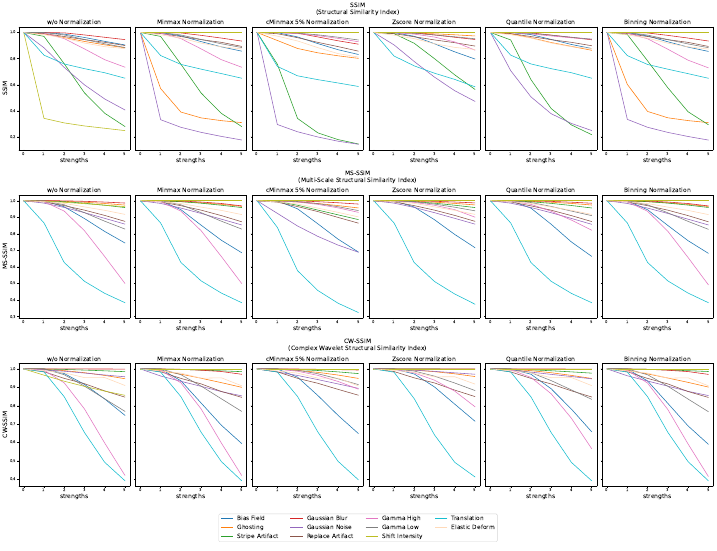}
\caption{Median scores of SSIM-based reference metrics SSIM (top), MS-SSIM (middle), and CW-SSIM (bottom) across 100 images, distorted with increasing strengths (0: reference, 1: hardly/not visibly distorted, 5: strongly distorted), grouped by kinds of distortions in different colors.}\label{fig:ssim}\label{fig:msssim}\label{fig:cwssim}
\end{figure}

\begin{figure}[htb]%
\centering
\includegraphics[width=\textwidth]{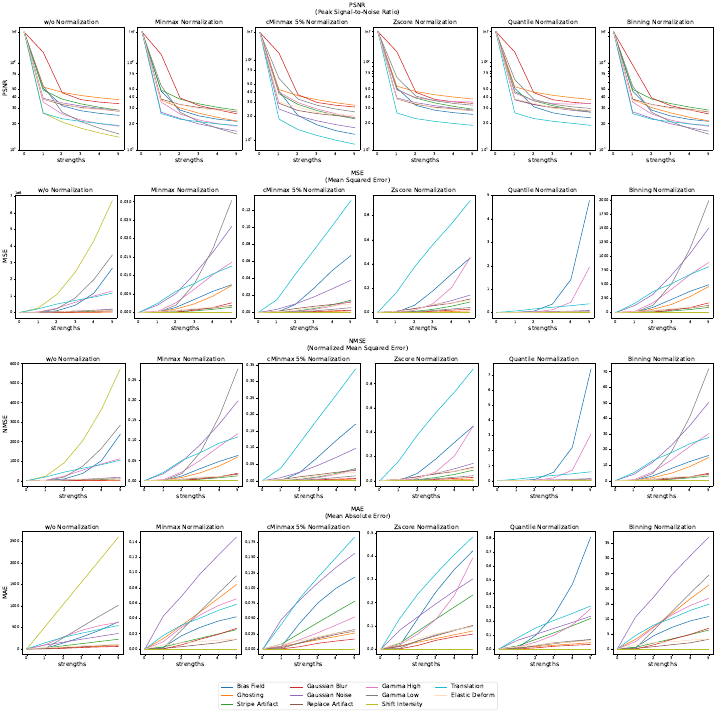}
\caption{Median scores of error-based reference metrics PSNR (top), MSE (second row), NMSE (third row) and MAE (bottom) across 100 images, distorted with increasing strengths (0: reference, 1: hardly/not visibly distorted, 5: strongly distorted), grouped by kinds of distortions in different colors.} \label{fig:mse}\label{fig:nmse}\label{fig:mae}\label{fig:psnr}
\end{figure}

\begin{figure}[h]%
\centering
\includegraphics[width=\textwidth]{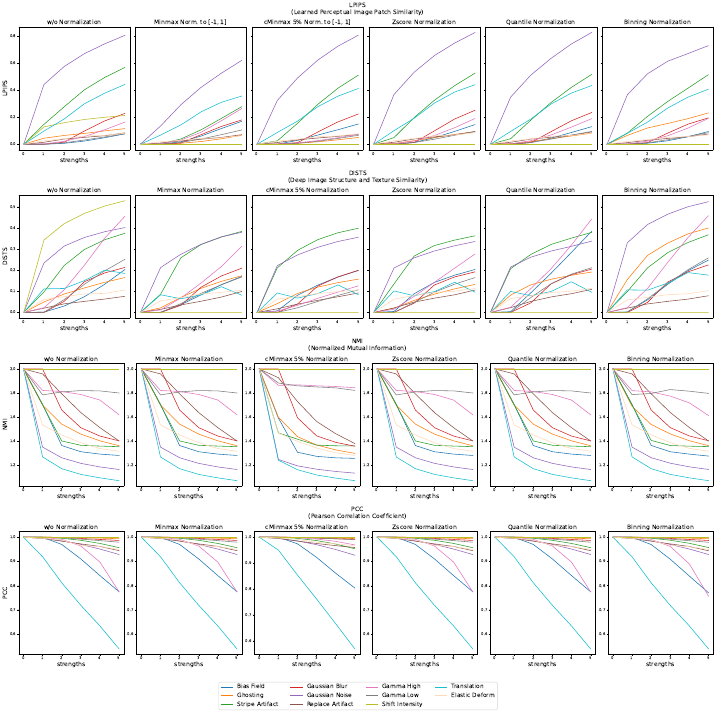}
\caption{Median scores of the learned reference metrics LPIPS (top) and DISTS (second row), and statistical dependency reference metrics NMI (third row) and PCC (bottom) across 100 images, distorted with increasing strengths (0: reference, 1: hardly/not visibly distorted, 5: strongly distorted), grouped by kinds of distortions in different colors. LPIPS and DISTS require normalized images with an intensity range around 0, therefore, analysis of these metrics was not performed without normalization.} \label{fig:lpips}\label{fig:dists}\label{fig:nmi}\label{fig:pcc}
\end{figure}

\clearpage
\subsection{Evaluation Plots for All Non-Reference Metrics and Normalization Methods}
\label{sec:nr_metrics_evaluation_plots}
In the following figures we present six  plots for each combination of reference metric, one for each normalization method, including w/o normalization. For each distortion and strength the median of all metric scores across all 100 cases is plotted, such that an increasing or decreasing trend is usually observed along the distortion strengths. Most non-reference metrics were designed to be applied to 8-bit integer images, which is the case after Binning normalization.

\begin{figure}[h]%
\centering
\includegraphics[width=\textwidth]{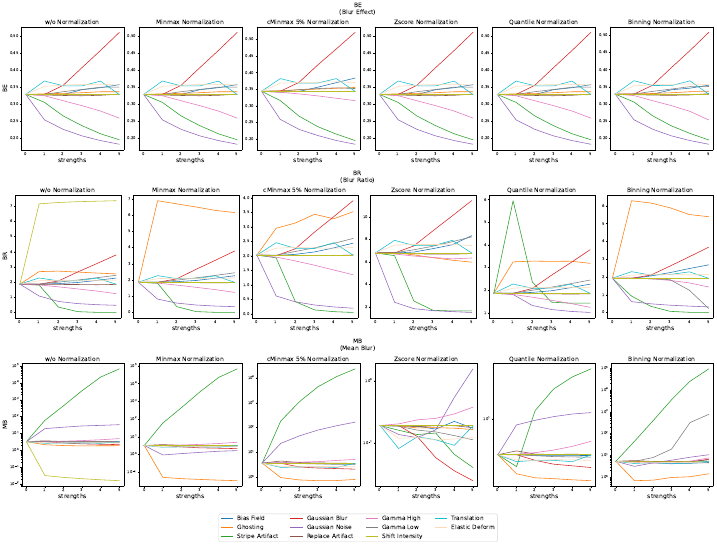}
\caption{Median scores of non-reference blurriness metrics BE (top), BR (middle) and MB (bottom) across 100 images, distorted with increasing strengths (0: reference, 1: hardly/not visibly distorted, 5: strongly distorted), grouped by kinds of distortions in different colors.} \label{fig:blur_effect}\label{fig:blur_ratio}\label{fig:mean_blur}
\end{figure}

\begin{figure}[h]%
\centering
\includegraphics[width=\textwidth]{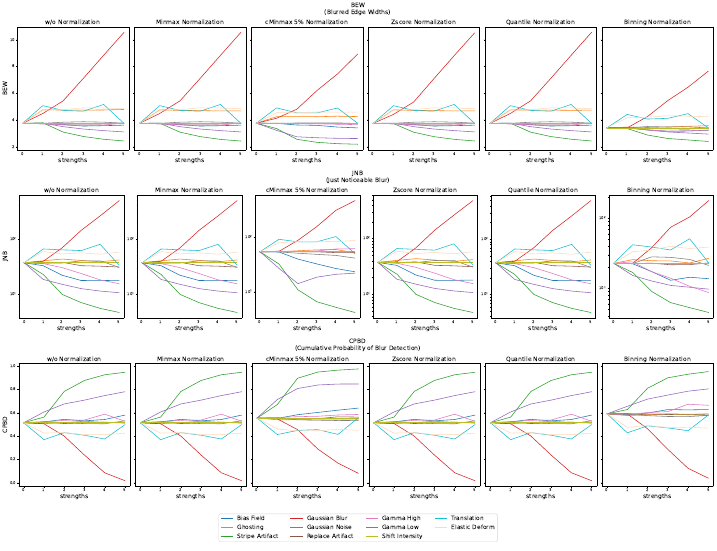}
\caption{Median scores of non-reference blurriness metrics BEW (top), JNB (middle) and CPBD (bottom) across 100 images, distorted with increasing strengths (0: reference, 1: hardly/not visibly distorted, 5: strongly distorted), grouped by kinds of distortions in different colors.} \label{fig:blurMarz}\label{fig:blurJNB}\label{fig:blurCPBD}
\end{figure}

\begin{figure}[h]%
\centering
\includegraphics[width=\textwidth]{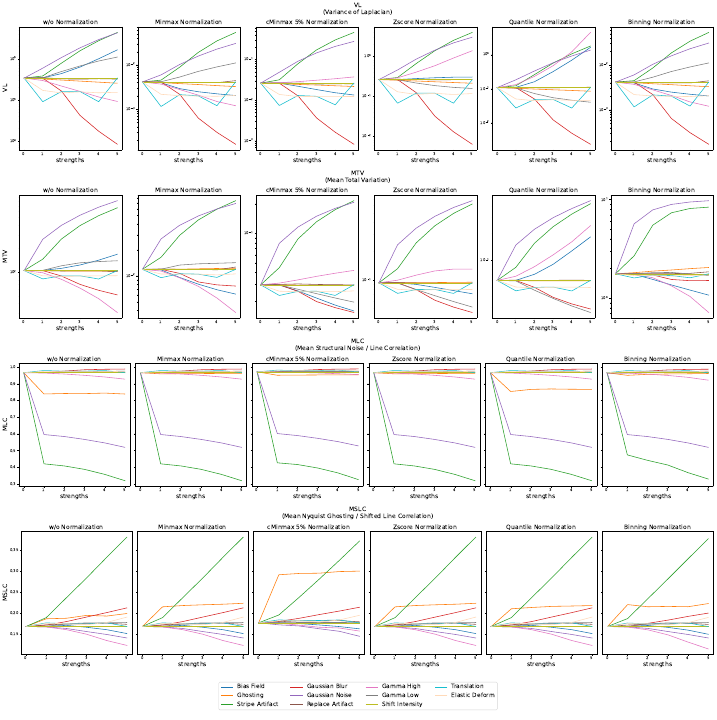}
\caption{Median scores of further non-reference blurriness, noisiness and MR acquisition quality metrics VP (top), MTV (second row), MLC (third row), and MSLC (bottom) metric across 100 images, distorted with increasing strengths (0: reference, 1: hardly/not visibly distorted, 5: strongly distorted), grouped by kinds of distortions in different colors.} \label{fig:mean_nyquist_ghosting}\label{fig:mean_struct_noise}\label{fig:var_laplace}, \label{fig:mean_total_var}
\end{figure}

\begin{figure}[h]%
\centering
\includegraphics[width=\textwidth]{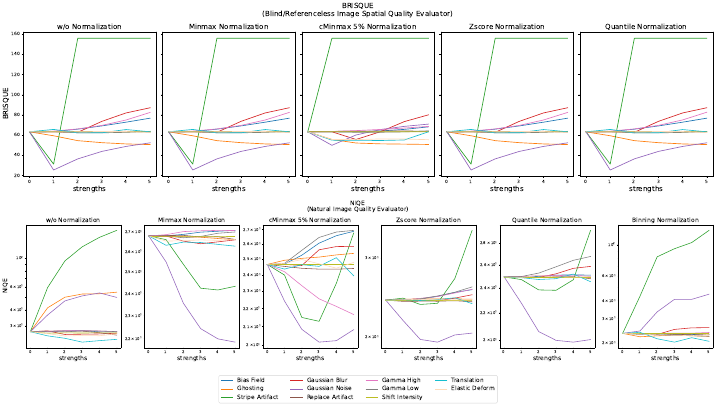}
\caption{Median scores of the learned non-reference quality metrics BRISQUE (top), and NIQE (bottom) metric across 100 images, distorted with increasing strengths (0: reference, 1: hardly/not visibly distorted, 5: strongly distorted), grouped by kinds of distortions in different colors.} \label{fig:brisque}\label{fig:niqe}
\end{figure}

\clearpage
\subsection{Evaluation Plots for Downstream Task Metric}
\label{sec:seg_metrics_evaluation_plots}
In the following figures we present six  plots for each combination of reference metric, one for each normalization method, inlcuding w/o normalization. For each distortion and strength the median of all metric scores across all 100 cases is plotted, such that an increasing or decreasing trend is usually observed along the distortion strengths.
\begin{figure}[h]%
\centering
\includegraphics[width=\textwidth]{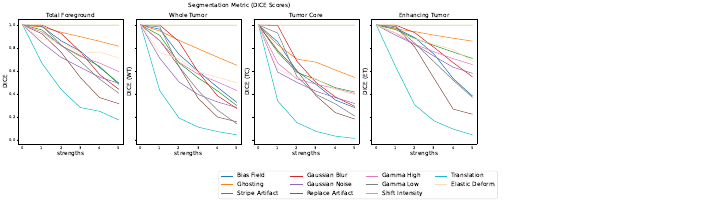}
\caption{Mean scores of the DSC segmentation metric across 100 segmentation pairs derived from a reference a distorted image with increasing strengths applied (0: reference, 1: hardly/not visibly distorted, 5: strongly distorted). The mean scores are grouped by kinds of distortions in different colors. The "Total Foreground" class includes the three disjoint classes "Whole Tumor" (mainly tumor surrounding edema),  "Tumor Core" (mainly necrotic areas) and "Enhancing Tumor" (vital tumor cells, taking up contrast media).  } \label{fig:dice}\label{fig:dice_label1}\label{fig:dice_label2}\label{fig:dice_label3}
\end{figure}
\end{document}